\documentclass[aps,prl,twocolumn,floatfix]{revtex4-2}
\usepackage{bm}
\usepackage{graphicx}
\usepackage{amsmath}
\usepackage{amssymb}
\usepackage[utf8]{inputenc}
\usepackage[T1]{fontenc}
\usepackage{color}
\usepackage{float}
\usepackage{cancel}
\usepackage{upgreek}
\usepackage{subfigure}
\usepackage[normalem]{ulem}
\usepackage[unicode=true,colorlinks=true,citecolor=blue]{hyperref}
\usepackage{placeins}
\usepackage{lipsum}
\usepackage{epsfig}
\usepackage{booktabs}
\usepackage[utf8]{inputenc}
\usepackage{wrapfig}
\newcommand{\nix}[1]{}

\usepackage{ulem}
\usepackage{physics}

\newcommand{\Ug}{U_{\rm g}^{\rm eff}}
\newcommand{\Upc}{U^{\rm circ}_{\rm pc}}

\newcommand{\sdg}[1]{\textcolor{blue}{#1}}

\begin{document}

\title{THz radiation induced circular Hall effect in graphene}

\author{S. Candussio$^1$, S. Bernreuter$^1$, T. Rockinger$^1$, K. Watanabe$^2$, T. Taniguchi$^3$, J. Eroms$^1$, I.A. Dmitriev$^{1,4}$, D. Weiss$^1$, and S.D. Ganichev$^{1,5}$
}

\affiliation{$^1$Terahertz Center, University of Regensburg, 93040 Regensburg, Germany}
\affiliation{$^2$Research Center for Functional Materials, National Institute of Material Science, 1-1 Namiki, Tsukuba 305-0044, Japan}
\affiliation{$^3$International Center for Materials Nanoarchitectonics, National Institute of Material Science, 1-1 Namiki, Tsukuba 305-0044, Japan}
\affiliation{$^4$Ioffe Institute, 194021 St. Petersburg, Russia}
\affiliation{$^5$ CENTERA, Institute of High Pressure Physics PAS, 01142 Warsaw, Poland}

\begin{abstract}
We report on the observation of the circular transversal terahertz photoconductivity in monolayer graphene supplied by a back gate. The photoconductivity response is caused by the free carrier absorption and reverses its sign upon switching the radiation helicity. The observed dc Hall effect manifests the time inversion symmetry breaking induced by circularly polarized terahertz radiation in the absence of a magnetic field. For low gate voltages, the photosignal is found to be proportional to the radiation intensity and can be ascribed to the alignment of electron momenta by the combined action of THz and static electric fields as well as by the dynamic heating and cooling of the electron gas. Strikingly, at high gate voltages, we observe that the linear-in-intensity Hall photoconductivity vanishes; the photoresponse at low intensities becomes superlinear and varies with the square of the radiation intensity. We attribute this behavior to the interplay of the second- and fourth-order effects in the radiation electric field which has not been addressed theoretically so far and requires additional studies.
\end{abstract}

\maketitle

\section{Introduction}

Optoelectronic phenomena in graphene, providing a highly effective means for the manipulation and control of carriers by radiation from the visible to terahertz (THz) frequency range, are subject of enormous current interest, see e.g.~\cite{Xia2009a,Bonaccorso2010,Liu2011,Echtermeyer2011,Koppens2011,Vicarelli2012,Engel2012,Grigorenko2012,Bao2012,Jariwala2013,Glazov2014,Koppens2014,Sun2014,Mueller2014,Sun2016,Sanctis2018,Wang2019,Tan2020}. In the last decade it was demonstrated that a circularly polarized radiation can produce a dc electric current whose direction and magnitude are controlled by the radiation helicity, see e.g.~\cite{Oka2009,Karch2010,Karch2011,Jiang2011,Kitagawa2011,Ivchenko2012,Glazov2014,Qian2018,Zhu2019,McIver2020,Sato2019,Matyushkin2020,Otteneder2020,Candussio2021,Candussio2021a,Durnev2021,Durnev2021a}. A particularly intriguing phenomenon is the circular Hall effect arising in the absence of static magnetic fields. 
The Hall current can appear either in unbiased samples, where it is driven by the crossed electric and magnetic fields of a circularly polarized wave and is termed the dynamic circular Hall effect~\cite{Karch2010,Jiang2011,Ivchenko2012,Glazov2014,Qian2018,Zhu2019}, or as circular transverse photoconductivity in the presence of a dc current, where it is driven solely by the electric field of the wave and is also termed the photovoltaic Hall effect~\cite{Oka2009,Kitagawa2011,McIver2020,Sato2019,Durnev2021a}.
At low intensity $I$ of illumination, the lowest-order transverse photoconductivity $\propto I$ has two contributions, one coming from the optical alignment of electron momenta and the other from the dynamic heating and cooling of the electron gas~\cite{Durnev2021a}. In addition, a high-intensity circularly polarized light can open gaps in the Dirac spectrum, which, as predicted in Ref.~\cite{Oka2009} and recently demonstrated by applying mid-infrared radiation in Ref.~\cite{McIver2020}, also leads to a photoinduced dc Hall current governed by the direct interband couplings. For highly doped samples and/or for terahertz radiation with relatively small photon energies not exceeding several meV, the spectrum reconstruction effects \cite{Oka2009} driven by the interband couplings should be less pronounced. In this case, the intraband free carrier absorption \cite{Durnev2021a} is expected to dominate the circular Hall effect in photoconductivity. 

Here we report an observation and study of the circular transverse photoconductivity in graphene induced by terahertz radiation of moderate and large intensity. We demonstrate that absorption of the terahertz radiation results in a Hall photocurrent whose direction changes to the opposite with inversion of either the bias voltage polarity or the radiation helicity. The detected Hall photoresponse is negligibly small at the charge neutrality point (CNP). At low gate voltages away from the CNP, the photoresponse is proportional to the radiation intensity $I$ at low intensities sometimes saturating at higher $I$. Such behavior is consistent with theory predicting the circular transverse photoconductivity of second order in the radiation electric field $\bm E$, see Refs.~\cite{Belinicher1981,Oka2009,Durnev2021a}.
Strikingly, at  high electron densities corresponding to higher gate voltages, the observed Hall photosignals exhibit a superlinear intensity dependence and vary as $I^2$. Such behavior is detected for several frequencies ranging from 0.78 to 3.33~THz. Unlike the transverse photoconductivity, the longitudinal photoconductivity signal measured parallel to the bias current was insensitive to the radiation helicity. We discuss the observed THz radiation-induced circular Hall effect in terms of intraband electron kinetics in the presence of static and circular high-frequency electric fields. The corresponding quasiclassical theory \cite{Durnev2021a} is consistent with our experimental results at low gate voltages and clarifies why the linear-in-$I$ contributions show up away from the CNP and can drop down at high gate voltages. An extension of this theory to higher orders is required to explain the emergence of strong quadratic-in-$I$ contributions to the circular Hall effect dominating the observed THz photosignals at high gate voltages and high intensities.

\section{Samples and methods}
\label{samples_methods}
\begin{figure}
	\centering
	\includegraphics[]{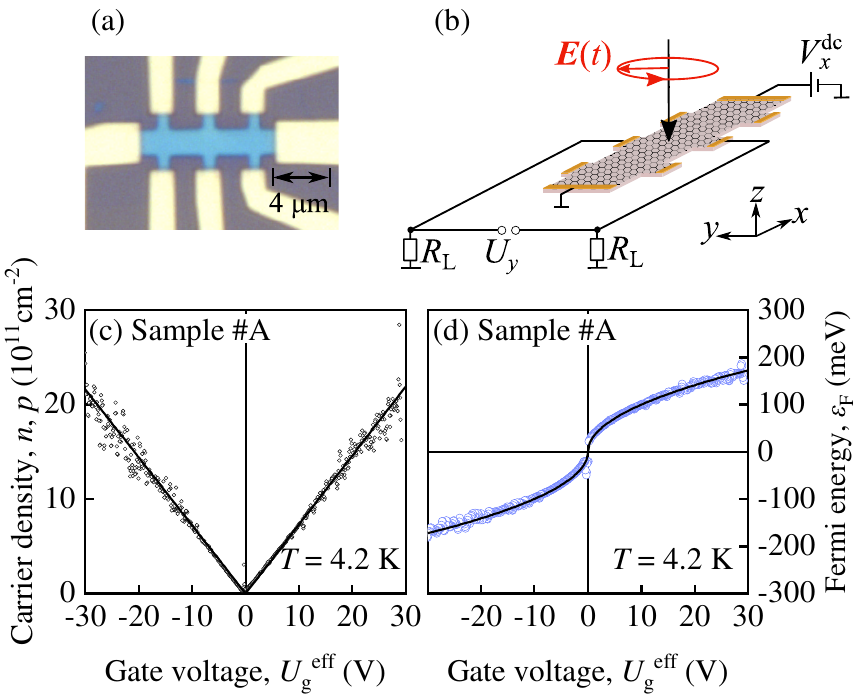}
	\caption{
	(a) A microscope image of the Hall bar structure (sample \#A). (b) The experimental setup. The sample is irradiated with circularly polarized light along $z$-direction at normal to the sample surface. A \textit{dc} bias voltage $V_x^\text{dc}$ is applied in $x$ direction along the Hall bar. The transversal photosignal $U_y$ is picked up as a voltage drop over load resistors $R_{\rm L} = 50 \,\Omega$. Subtracting the photosignal for opposite $V_x^\text{dc}$ polarities yields the photoconductivity signal $U_{pc}=[U_y(V_x^\text{dc})-U_y(-V_x^\text{dc})]/2$.
	Bottom panels show characteristics of the investigated graphene sample obtained from magneto-transport measurements.  (c) The carrier density  versus the effective gate voltage $U_g^{\rm eff} = U_g -U_g^{\rm CNP}$. The solid line shows a linear fit after $n,p\, [{\rm cm}^{-2}]= 0.75\times 10^{11}\,|U_g^{\rm eff}|[V]$.  (d) The Fermi energy $\varepsilon_F$ (blue dots) determined from the carrier density (c). The solid black line corresponds to the linear fit from panel (c).}
	\label{transport}
\end{figure}

 The  exfoliated graphene/hexagonal boron nitride stacks~\cite{Dean2010,Wang2013,Sandner2015} were prepared as Hall bar structures, see the microscope picture of sample \#A in Fig.~\ref{transport}~(a). All investigated samples \#A, \#B, and \#C had a Hall bar width of $2\,\upmu$m and length of $L=9\,\upmu$m and manifested similar transport- and photoresponce. By changing the back gate voltage the carrier density could be tuned in a wide range and varied symmetrically with the effective gate voltage, $U_g^{\rm eff}$, as $n,p\, [{\rm cm}^{-2}]= 0.75\times 10^{11}\,|U_g^{\rm eff}|[V]$, where $n$ and $p$ are electron and hole sheet densities at positive and negative $U_g^{\rm eff}$, respectively. Here $U_g^{\rm eff}=U_g - U_g^{\rm CNP}$ was determined separately for every cool down tracing slight shifts of the gate voltage $U_g^{\rm CNP}$ corresponding to the charge neutrality point (CNP). Figure \ref{transport}\,(c) depicts the carrier density obtained from classical Hall measurements at liquid helium temperature in the absence of THz illumination. For these measurements an \textit{ac} current of 10~nA at a frequency of $12\,$Hz was applied to the sample. The corresponding Fermi level position $\varepsilon_\text{F}$ as a function of $U_g^{\rm eff}$ is presented in Fig.~\ref{transport}\,(d). It was calculated using the conventional relations $\varepsilon_\text{F}=\hbar v_\text{F} \sqrt{\pi n}$, $\varepsilon_\text{F}=- \hbar v_\text{F} \sqrt{\pi p}$ and a
 standard value $v_F=10^6$~cm/s for the electron velocity in graphene.

For studies of the circular Hall effect we used a high power pulsed THz molecular gas laser~\cite{Shalygin2006,Plank2016,Dantscher2017} pumped by a transversely excited atmospheric pressure (TEA) CO$_2$ laser~\cite{Ganichev2003}. As illustrated in Fig.~\ref{transport}~(b) the sample was illuminated with circularly polarized THz pulses at normal incidence. The laser operated at frequencies $f= 0.78, 2.02,$
and $3.33\,\text{THz}$ with pulse duration of about 100 ns and repetition rate of 1 Hz. The laser power, analyzed by photon drag detectors, was of the order of tens of kW varying for different frequencies. Taking into account the diameter of the Gaussian beam profile (1.5-3~mm), known from measurements using a pyroelectric camera, intensities up to 150 kW/cm$^2$ could be achieved on the sample position. To control the laser radiation intensity arriving on the sample we placed two grid polarizers into the optical path, where the first one was rotatable and the second one was at a fixed position~\sdg{~\cite{Hubmann2019,Candussio2021a}.} The radiation polarization was modified by placing $\lambda/4$-plates in front of the sample, which were rotated by an angle $\varphi$ between the $c-$axis of the plate and the electric field vector of the laser radiation after the second grid polarizer. In most of the experiments we used angles $\varphi=45^\circ$ and $\varphi=135^\circ$ corresponding to the right handed ($\sigma^+$) and  left handed ($\sigma^-$) circularly polarized light, respectively. The sample was placed into an optical temperature-regulated continuous flow cryostat where it was cooled down to $T = 4.2\,$K. The $z$-cut crystal quartz windows were covered by a black polyethylene film transparent for THz radiation preventing the undesired illumination of the sample with room light.

To change the sample conductivity under THz illumination a \textit{dc} bias voltage $V_x^\text{dc}$ was applied between source and drain contacts, i.e. along the long side of the Hall bar ($x-$direction)~\footnote{Note that few final measurements were carried out for $V_y^\text{dc}$ applied between contacts neighboring the damaged source and drain contacts, which had no apparent effect on the signal recorded for the Hall voltage induced in the middle of the Hall bar.}.
The transverse (Hall) photosignal was picked up between two oppositely placed contacts at the middle of the Hall bar and measured as a voltage drop $U_y(V_x^\text{dc})$ over load resistors $R_L=50\, \Omega$, see Fig.~\ref{transport}(b). 
For comparison, we also studied the longitudinal photoconductivity within the two-terminal measurement scheme in which a voltage drop $U_y(V_y^\text{dc})$ or $U_x(V_x^\text{dc})$ was measured between two contacts biased over a load resistor either across the Hall bar (Fig.~\ref{bias} in the main text) or along the Hall bar (Fig.~\ref{intdep_oldsample} in the supplementary materials).

\section{Results}

\begin{figure}
	\centering
	\includegraphics[]{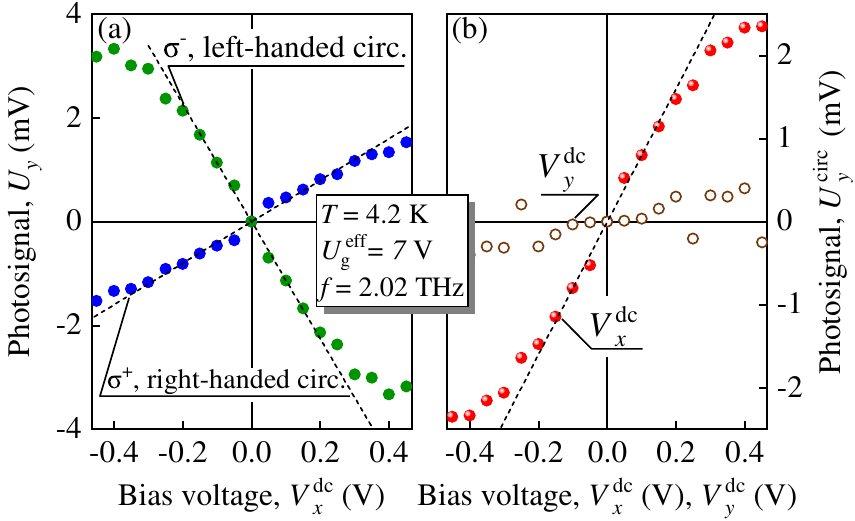}
	\caption{(a) Transversal photosignal picked up in sample \#A perpendicular to the direction of the applied {\it dc} bias voltage $V_{x}^\text{dc}$ which is swept at a fixed gate voltage of $U_{\rm g}^{\rm eff}=7$~V for right- and left-handed circularly polarized light. (b) The calculated helicity dependent parts $U_y^\text{circ}=(U_y^{\sigma^+}-U_y^{\sigma^-})/2$ of the Hall signal (\textit{dc} bias voltage $V_x^\text{dc}$ along $x$, full dots) and of the longitudinal signal (\textit{dc} bias voltage $V_y^\text{dc}$ along $y$, open dots). Dashed linear fits are a guide for the eye.
	}
	\label{bias}
\end{figure}

Measurements under circularly polarized radiation revealed the helicity-sensitive Hall photoresponse. It was detected in $y$ direction across the Hall bar perpendicular to the direction of the applied bias voltage $V_x^\text{dc}$. Figure~\ref{bias}(a) shows a typical bias voltage dependence of the photoresponse to right-handed ($\sigma^{+}$) and left-handed ($\sigma^{-}$) circularly polarized radiation. It demonstrates that for fixed radiation helicity the signal is proportional to the applied bias voltage. It vanishes at zero bias voltage, and changes the sign by switching from negative to positive $V_x^\text{dc}$. These facts reveal that the main signal comes from the change of the sample conductivity. We also observed that the variation of the signal with bias voltage remained linear for $V_x^\text{dc}$ up to $\pm 0.2$\,V, with a tendency to saturate at higher bias voltages, see Fig.~\ref{bias}(a).

\begin{figure}
	\centering
		\includegraphics[width=\linewidth]{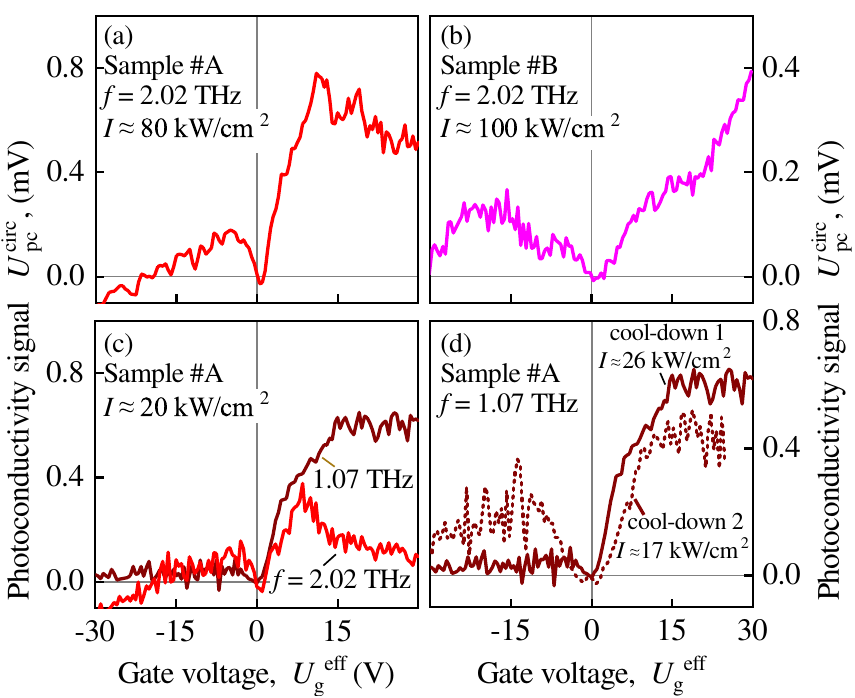}
	\caption{The transversal photoconductivity $\Upc$ excited by circularly polarized radiation, extracted using Eqs.~(\ref{eqcirc}) and (\ref{eqodd}) from $U_y$ measured by sweeping the gate voltage $\Ug$ from $-30$ to $+30$\,V. (a) Gate voltage dependence of $\Upc$ measured in sample \#A at a frequency of $f= 2.02$\,THz and with an intensity of $I\approx 80 \textrm{kW/cm}^2$. (b) $\Upc$ detected in sample \#B at the same frequency. (c) $\Upc$ in sample \#A as a function of $\Ug$ for two different frequencies $f=2.02$ and $f=1.07$\,THz at $I\approx 20\,\mathrm{kW/cm}^2$. (d) $\Upc$ detected after two different sample cool downs on sample \#A under $f=1.07$\,THz illumination. $\Upc$  after Eqs.~(\ref{eqcirc}) and (\ref{eqodd}) was calculated using $T = 4.2$\,K measurements at $V_x^\text{dc}=\pm 0.2$~V for sample \#A and $V_x^\text{dc}=\pm 0.1$~V for sample \#B.}	
	\label{frequency}
\end{figure}

Importantly, an inversion of the radiation helicity consistently inverted the sign of the Hall photoconductivity signal, see Fig.~\ref{bias}(a). Rotating the $\lambda/4$-plate we observed that the signal closely follows the degree of circular polarization ${P_\text{circ}=(I^{\sigma^+}-I^{\sigma^-})/(I^{\sigma^+}+I^{\sigma^-})}$ so that $U_y \propto P_{\rm circ}$ (not shown). Here, $I^{\sigma^+}$ and $I^{\sigma^-}$ are intensities of the right- and left-handed circularly polarized radiation. We emphasize that such a helicity sensitive behaviour was only observed for the Hall photoconductivity, i.e. for $U_y$ detected with the {\it dc} bias $V_x^\text{dc}$ applied along the Hall bar. For longitudinal photoconductivity, in contrast, the signals for right and left handed circular polarization were almost identical, see Fig.~\ref{intdep_oldsample} in the supplementary materials. To explore the functional behavior of the helicity dependent signal, we write it as
\begin{equation}
\label{eqcirc}
	U_y^{\rm circ}=\frac{U_y^{\sigma^+}	-U_y^{\sigma^-}	}{2},
\end{equation}
where $U_y^{\sigma^+}$ and $U_y^{\sigma^-}$ are the photosignals generated by right- and left-handed circularly polarisation, respectively. The variation of the Hall circular photoresponse with  bias voltage is shown in Fig.~\ref{bias}(b) together with the results of the longitudinal circular photoresponse obtained for the bias voltage applied in $y$ direction. The circular longitudinal signal is almost zero, see Fig.~\ref{bias}(b) and Fig.~\ref{intdep_oldsample} in the supplementary materials, and, similar to the Hall signal, changes its sign upon inversion of the bias voltage polarity.

While the sign change upon switching the bias voltage polarity was detected in all measurements, the magnitude of signals for $V_x^\text{dc}$ and $-V_x^\text{dc}$ were slightly different, see Fig.~\ref{bias}(a) and Fig.~\ref{current1} in the supplementary materials. This observation is attributed to the generation of the photogalvanic currents~\cite{Glazov2014,Candussio2021a}. Using that, by definition, the linear-in-$V_x^\text{dc}$ photoconductivity signal should have opposite sign for positive and negative bias voltages $V_x^\text{dc}$ whereas the the photogalvanic current should be insensitive to the polarity of $V_x^\text{dc}$, we extracted the photoconductivity contribution $U_{\rm pc}^{\rm circ}$ as an odd part of $U_y^{\rm circ}(V_x^\text{dc})$,
\begin{equation}
\label{eqodd}
U_{\rm pc}^{\rm circ}= \frac{U_y^{\rm circ}(V_x^\text{dc})-U_y^{\rm circ}(-V_x^\text{dc}) }{2}.
\end{equation}

Figure~\ref{frequency} illustrates the gate voltage dependence of the corresponding circular photoconductivity signal $U_{\rm pc}^{\rm circ}$. Panels (a) and (b) show $U_{\rm pc}^{\rm circ}$ obtained in samples \#A and \#B under intense $f=2.02$\,THz radiation with $I\approx 100$~kW/cm$^{-2}$. These traces demonstrate that the circular photoconductivity is negligible at the CNP, and, for small gate voltages, increases almost linearly and symmetrically with $\Ug$. At higher gate voltages, however, the dependences become asymmetric: for positive gate voltages $U_{\rm pc}^{\rm circ}$ increases further and typically saturates at high gate voltages, whereas for the negative $\Ug$ the Hall photoconductivity decreases and may even change the sign. Similar results are obtained for lower intensities and other frequencies, see Fig.~\ref{frequency}(c). 
The electron-hole asymmetry of the circular photoconductivity was detected in all measurements, but was different for different cool downs. Figure~\ref{frequency}(d) shows an exemplary gate voltage dependence of the signal obtained for two different cool downs. While for positive gate voltages both amplitude and functional behavior are similar, for negative gate voltages in one of the cool downs the signal is close to zero for all $\Ug$. Therefore, in the following we focus on the data obtained for positive gate voltages. 

\begin{figure}
	\centering
	\includegraphics[]{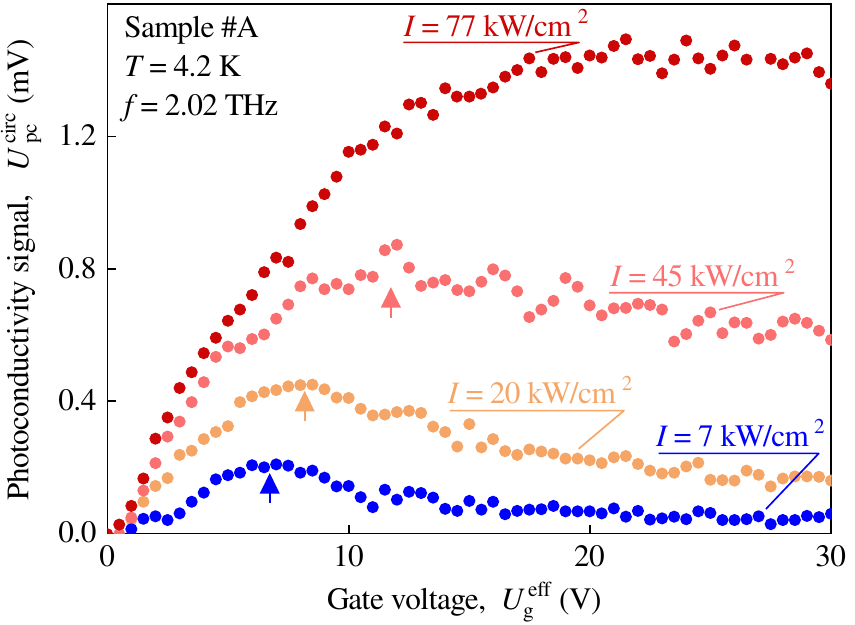}
	\caption{Gate voltage dependences of the helicity dependent part of the Hall photoconductivity signal $U^{\rm circ}_{\rm pc}$ for four different intensities $I$ as indicated. Up arrows point to the maxima of $U^{\rm circ}_{\rm pc}(\Ug)$ . These data were obtained on sample \#A at $T=4.2$\,K and $V_x^\text{dc}=\pm 0.2$ V.}
	\label{gatedep}
\end{figure}

Figure~\ref{gatedep} shows the gate voltage dependences obtained for $f=2.02$~THz with different radiation intensities. It demonstrates that the circular photoconductivity depends non-monotonically on $\Ug$: at low intensities it increases almost linearly with raising  $\Ug$, reaches a maximum and decreases for large $\Ug$. The maximum position depends on the radiation intensity and shifts to higher $\Ug$ with increasing radiation intensity. 

\begin{figure}
	\centering
	\includegraphics[width=\linewidth]{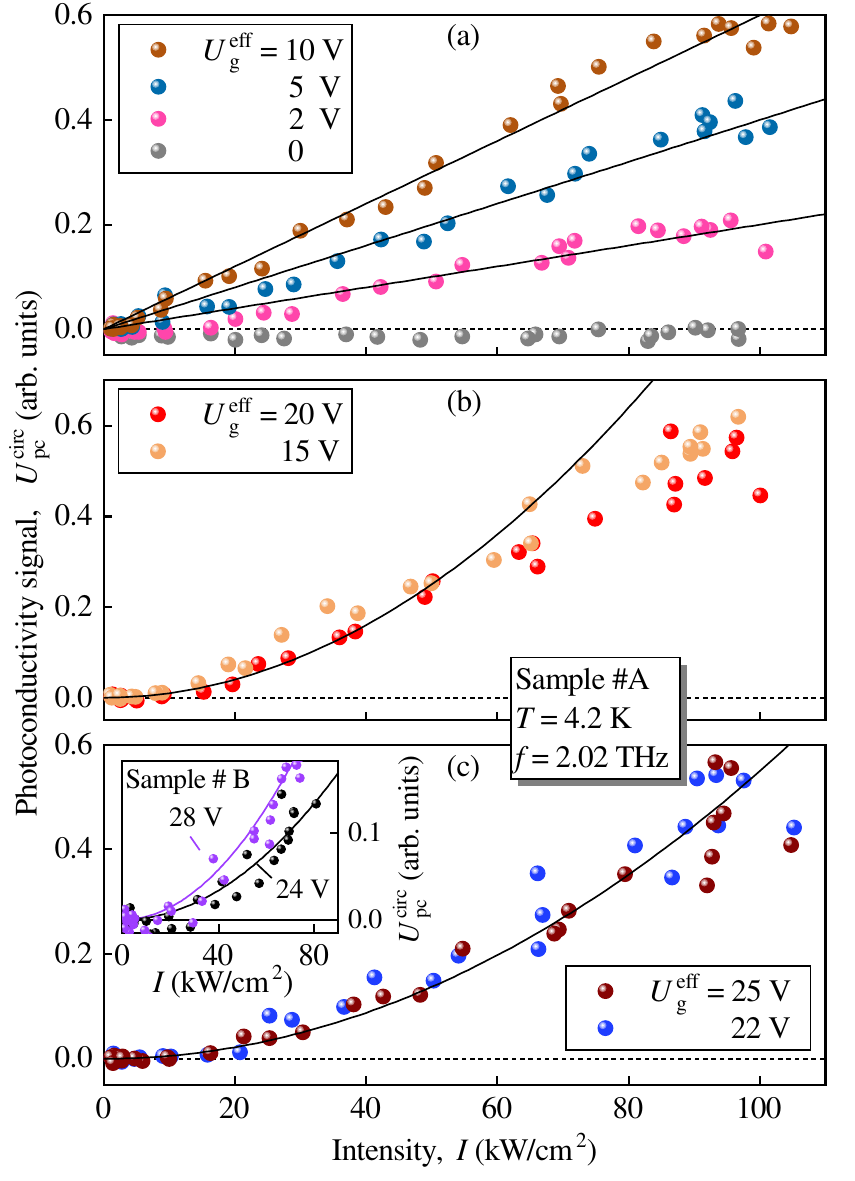}
	\caption{ The intensity dependences of $\Upc$ for different effective gate voltages $\Ug$ measured on sample \#A under $f=2.02$\,THz excitation and $V_x^\text{dc}=\pm 0.2$ V dc bias. Panel (a) presents the data for low $\Ug$; solid lines are linear fits after $\Upc = a  I$ yielding $a=2$, 4 and 6 $\upmu$V cm$^2$/kW. Solid lines in panels (b) and (c) are fits after $\Upc = A  I^2$. The inset presents the intensity dependences at $\Ug = 24$ and $28$\,V measured on sample \#B under $f=2.02$\,THz excitation and $V_x^\text{dc}=\pm 0.1$ V dc bias. The obtained coefficients are $A=0.1$ (b), 0.055 (c), 0.021 and 0.036 (inset), in units of $\upmu$V cm$^4$/kW$^2$.
}
	\label{intdep}
\end{figure}

\begin{figure}
	\centering
	\includegraphics[width=\linewidth]{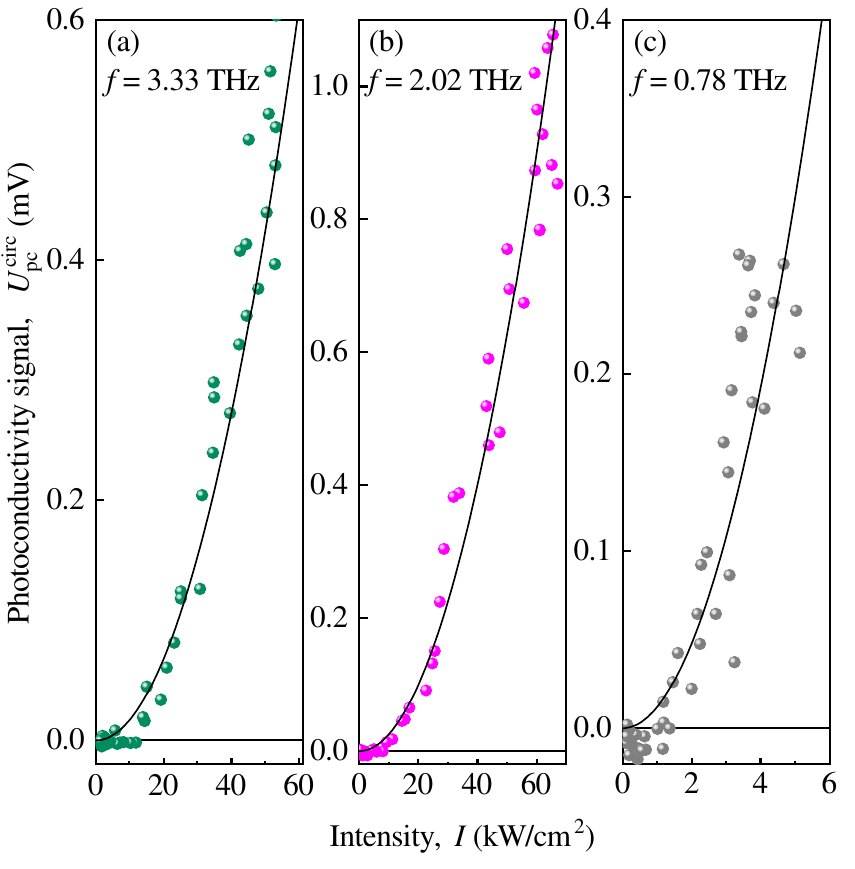}
	\caption{The intensity dependence of $\Upc$  obtained on sample \#A at high gate voltage, $\Ug = 30$\,V, and three different frequencies: (a) $f= 3.33$\, THz; (b) $f=2.02$\, THz, and (c) $f= 0.78$\, THz at $V_y^\text{dc}=\pm 0.2$ V. Solid lines are fits after $\Upc = A \, I^2$ with $A$ = 0.17, 0.25, and 12 $\upmu$V cm$^4$/kW$^2$ in panels (a), (b), and (c), respectively.}
	\label{frequency2}
\end{figure}

While for low gate voltages the circular photoconductivity linearly increases with increasing radiation intensity $I$, see Fig.~\ref{intdep}, or saturates at high $I$ (not shown), at high gate voltages it becomes superlinear.  In the latter case, in particular, for the highest gate voltages, the data can be well fitted by $\Upc = A(f) \times I^2$, where $A(f)$ is a fit parameter, see Fig.~\ref{intdep}(b) and (c) for sample \#A and the inset in  Fig.~\ref{intdep}(c) for sample \#B. The superlinearity at high gate voltages has been detected for all radiation frequencies, see Fig.~\ref{frequency2}. The coefficients $A(f)$ increase at low frequency, for instance, $A(0.78\,\textrm{THz})/A(3.33\,\textrm{THz})\approx 70$.  Note that in our laser systems the highest available intensities become lower at low radiation frequencies, see Fig.~\ref{frequency2}.

\section{Discussion}

We begin the discussion of the observed circular photoconductivity by reviewing the results of a phenomenological theory based on general symmetry arguments. When an isotropic system is excited by normally incident spatially homogeneous terahertz radiation, the minimal order {\it dc} photocurrent, linear with respect to the {\it dc} electric field $E^{\rm{dc}}_x$ applied in  $x$ direction and quadratic in the radiation electric field $\bm E \exp(-i\omega t)+\bm E^* \exp(i\omega t)$, is fully characterised \cite{Belinicher1981} by three transport coefficients $\gamma_k$,
\begin{align}
\label{xxgen}
j_x &= (\gamma_1 + \gamma_2 S_1)|\bm E|^2 E^{\rm{dc}}_x\,,\\
j_y &=  (\gamma_2 S_2 + \gamma_3 S_3)|\bm E|^2 E^{\rm{dc}}_x\,,
\label{yxgen}
\end{align}
where $S_1=(E_x^2-E_y^2)/|\bm E|^2$, $S_2=(E_xE^*_y+E_x^*E_y)/|\bm E|^2$, and $S_3=i(E_xE^*_y-E_x^*E_y)/|\bm E|^2$ are Stokes parameters describing the polarization of the radiation field. In the case of circular polarization, the anisotropic terms proportional to $S_1=S_2=0$ vanish, and the photoresponce is reduced to
\begin{align}
\label{xx}
j_x &= \gamma_1 |\bm E|^2 E^{\rm{dc}}_x\,,\\
j_y &=  \eta \gamma_3 |\bm E|^2 E^{\rm{dc}}_x\,,
\label{yx}
\end{align}
where the helicity $\eta=\pm 1$ represents the two possible values of $S_3=\pm 1$ for right- and left-handed circular polarization. The form of the photoresponse given by Eq.~(\ref{yx}) is analogous to the conventional Hall effect, with the circularly polarized radiation playing the role of the time reversal symmetry breaking field instead of magnetic field and with the diagonal and Hall components of the photoconductivity tensor being even and odd with respect to the helicity $\eta$, in accord with our observations, see Figs.~\ref{bias}, \ref{frequency}, \ref{gatedep}, and \ref{frequency2} for the Hall photocurrent, and Figs.~\ref{bias}(b) and \ref{intdep_oldsample} for the longitudinal photocurrent. Note that in Fig.~\ref{bias}(b) the longitudinal current was detected by applying the bias voltage across the Hall bar, along $y$ direction, and by measuring the corresponding $j_y$ component of the photocurrent. 

In general, the photoconductivity response to THz radiation may be caused by either direct interband or indirect intraband optical transitions, the latter being due to the scattering-assisted free carrier absorption. In degenerate graphene at low temperatures, the interband processes require photon energies larger than twice the Fermi energy since the final states for optical transition should be free. Thus the interband photoconductivity response should be maximal for gate voltages in the vicinity of the CNP and should vanish at large $\Ug$. The fact that in our experiments the signal vanishes at the CNP (see Figs.~\ref{bias}, \ref{frequency}, and \ref{gatedep}) demonstrates that direct inter-band optical transitions do not contribute to the observed photoresponce. In addition, in the presence of a high intensity radiation the interband optical coupling can lead to significant modification of the excitation spectrum of graphene \cite{Oka2009}. The corresponding enhanced photoresponce in the vicinity of spectral gaps of the dressed states is also unavailable in our study with THz photon energies much smaller than typical Fermi energies. Therefore, in the following we focus on the scattering-assisted intraband mechanisms of the photoresponse. 

For low gate voltages ($< 10~V$) we observed that $U_{\rm pc}^{\rm circ}$ increases linearly with $\Ug$ and is almost symmetric for positive and negative gate voltages, Fig.~\ref{frequency}(a)-(c) \footnote{Note that the symmetry in respect to the CNP has been found to be sensitive to the cool-down procedure. In experiments differing by the cool-down circle only we observed that in several cool downs the signal for negative $\Ug$ was almost absent, see Fig.~\ref{frequency}(d). This fact together with the mentioned cool-down dependent shift of the CNP indicates that the surface charge can be different in different measurements and may play an important role in the photoconductivity response.}. Importantly, for low gate voltages we also detected that the signal grows linearly with the radiation intensity, i.e., in agreement with Eq.~\ref{yx}, is proportional to the square of the radiation electric field $|\bm E|^2$, see Fig.~\ref{intdep}(a). 

The kinetic theory of the lowest-order transverse photoconductivity in two-dimensional materials with arbitrary dispersion was recently developed in Ref.~\onlinecite{Durnev2021a}. Within the semiclassical kinetic approach based on Boltzmann equation it has been shown that the circular Hall photocurrent, $j_y=\eta \gamma_3 |\bm E|^2 E^{\rm{dc}}_x$, see Eq.~(\ref{yx}), contains two contributions. One of them is caused by the optical alignment of electron momenta and the second is due to the dynamic heating and cooling of the electron gas. 
These contributions are associated with the excitation of the second (optical alignment) and zeroth (dynamic heating) angular harmonics of the Boltzmann distribution function, both oscillating in time with the radiation frequency. Both harmonics appear at the second perturbation order from the equilibrium Fermi distribution via successive perturbation by the THz and static electric fields. The magnitude of these perturbations is controlled by the corresponding dynamic relaxation rates $\tau_{n\omega}^{-1}=\tau_n^{-1}-i\omega$, where $\tau_n^{-1}$ denotes the relaxation rate for the corresponding $n$th static angular harmonics.
The explicit result of Ref.~\onlinecite{Durnev2021a} for $\gamma_3$ in the case of graphene, with the linear dispersion $\varepsilon= v p$, reads
\begin{align}
\nonumber
 \gamma_3 &= \sigma_0 e^2 v^2 \mathrm{Im}\left\{
 \alpha_\omega\tau_{0\omega}\left[
 \dfrac{\tau_1}{\varepsilon} +\dfrac{\varepsilon}{2}\left(\dfrac{\tau_1}{\varepsilon}\right)^\prime
 \right]^\prime\right.\\
&\left.- \dfrac{\alpha_\omega\varepsilon^2}{2}\left[
 \dfrac{\tau_{2\omega}}{\varepsilon}\left(\dfrac{\tau_1}{\varepsilon}\right)^\prime \right]^\prime
 -2\alpha_\omega\tau_{2\omega}\left(\dfrac{\tau_1}{\varepsilon}\right)^\prime\right\}_{\varepsilon=\varepsilon_F}\,.
\label{durnev16}
\end{align}
Here $\sigma_0=e^2\varepsilon_F \tau_1/\pi\hbar^2$ is the static conductivity, $e$ the elementary charge, $\varepsilon_F$ the Fermi energy, $\alpha_\omega=1+(1-i\omega\tau_1)^{-1}$, and primes denote derivatives with respect to kinetic energy $\varepsilon$ taken at the Fermi surface $\varepsilon=\varepsilon_F$. 

It is seen that the circular photoconductivity $\eta \gamma_3 |\bm E|^2$ is sensitive to the microscopic nature of scattering, which determines the energy dependence of the scattering rates $\tau_1^{-1}(\varepsilon)$ and $\tau_2^{-1}(\varepsilon)$ in the vicinity of the Fermi surface $\varepsilon=\varepsilon_F$ \footnote{In Ref.~\onlinecite{Durnev2021a} the energy relaxation rate $\tau_0^{-1}$ is considered to be $\varepsilon$-independent.}. In particular, one immediately observes that $\gamma_3$ vanishes in the important case $\tau_1=2\tau_2\propto\varepsilon$. This model represents scattering at the Coulomb centers relevant to graphene at low carrier densities, and is consistent with $\gamma_3=0$ at the CNP observed in our experiments. 

On the other hand, since the relevant scattering times $\tau_0\gg\tau_1\sim\tau_2\sim 1$~ps are much longer than $1/\omega$ for the relevant THz frequencies and low temperatures, we are mostly interested in the high-frequency regime $\omega\tau_n\gg 1$ of Eq.~(\ref{durnev16}). Interestingly, in this limit the dynamic heating ($\propto \tau_{0\omega}$) and optical alignment ($\propto \tau_{2\omega}$) terms in  Eq.~(\ref{durnev16}) cancel each other in the leading order which can be obtained by setting $\alpha_\omega=1$ and $\tau_{0\omega},\tau_{2\omega}=i\omega^{-1}$. As a result of this cancellation, in the high-frequency regime $\gamma_3$ scales as $\omega^{-3}$ unlike individual contributions scaling as $\omega^{-1}$. In particular, for the short-range scattering with $\tau_1=2\tau_2\propto\varepsilon^{-1}$, the high-frequency limit of Eq.~(\ref{durnev16}) reduces to $\gamma_3=-6e^4v^2/\pi\hbar^2\omega^3\varepsilon_F$. For the model $\tau_1=2\tau_2\propto\varepsilon/(\varepsilon^2+\varepsilon_0^2)$ combining the short-range and Coulomb scattering, 
 the theory of Ref.~\onlinecite{Durnev2021a} thus predicts the Hall photoresponse $\gamma_3$ which is maximized at some intermediate carrier density, corresponding to $\varepsilon_F\sim\varepsilon_0$, and decreases both towards the CNP and towards higher carrier densities.

Qualitatively, the observed circular photoconductivity at low gate voltages can be well described within the above mechanisms: the Hall photoconductivity signal is reversed with the change of radiation helicity, scales linearly with the square of the radiation electric field, see Fig.~\ref{intdep}(a), is zero at the CNP and almost symmetrically increases with increasing electron or hole density, see Figs.~\ref{frequency} and \ref{gatedep}. Moreover, the model described above is consistent with the data obtained for low radiation intensities and high carrier densities.
Indeed, Fig.~\ref{gatedep} demonstrates that for low radiation intensities the signal magnitude increases with increasing $\Ug$, approaches a maximum, and drops down at higher gate voltages. 

While at low intensities the circular photoconductivity signal strongly decreases at large carrier densities, an increase of the radiation intensity qualitatively changes the gate voltage dependence, see Fig.~\ref{gatedep}. The most essential modification, however, is that the intensity dependence of the circular photoconductivity at large carrier densities can no longer be described by Eqs.~(\ref{yx}) and (\ref{durnev16}), as the measured signal $U_{\rm pc}^{\rm circ}$ scales as $I^2\propto |\bm E|^4$, see Fig.~\ref{intdep}(b),(c) and \ref{frequency2}, while the theory developed so far is limited to effects of the minimal order, $U_{\rm pc}^{\rm circ}\propto I\propto |\bm E|^2$. Within the semiclassical approach of Ref.~\onlinecite{Durnev2021a}, our findings thus require calculation of the higher-order terms in the expansion 
\begin{equation}
\label{yx4}
j_y =  \eta\left(\gamma_3^{(2)} |\bm E|^2+\gamma_3^{(4)} |\bm E|^4 +\cdots\right) E^{\rm{dc}}_x\,,
\end{equation}
which should involve excitation of a larger number of different time and angular harmonics of the distribution function, and result in different combinations of the scattering rates and their derivatives at the Fermi surface. Such theory should explain the dominance of  
the $\gamma_3^{(4)} |\bm E|^2$ over $\gamma_3^{(2)}$ at increasingly lower intensity at higher gate voltages, consistent with the associated up-shift of $\Ug$ corresponding to the maximal photoresponse at higher intensities, see. Fig.\ref{gatedep}.

The analysis of Eq.~(\ref{durnev16}) above suggests that, in the high-frequency THz regime, contributions to $\gamma_3^{(4)}$ involving static angular harmonics that possess slower decay, $\tau_n^{-1}\ll\omega$, may play the most prominent role in the photoresponse. Especially important can be effects involving zeroth static angular harmonic which describe the radiation-induced changes in the energy distribution of carriers and are also frequently called heating effects. Indeed, at low temperatures this harmonic possesses the slowest decay, $\tau_0^{-1}\ll\tau_n^{-1}$, $n\neq 0$. In our experiments, a substantial electron heating is confirmed by the observation of polarization-independent longitudinal photoconductivity, see Fig.~\ref{intdep_oldsample}, which is caused by a decrease of electron mobility due to the heating effects (negative photoconductivity). Negative photoconductivity in degenerate systems is usually associated with enhanced momentum relaxation of hot electrons due to scattering processes involving acoustic phonons which yields a negative addition to $\tau_1$ growing with the radiation intensity. Similarly, heating effects can enter the circular photoconductivity (\ref{durnev16}) via  modified intensity-dependent scattering rates $\tau_n^{-1}(I)$. Due to their different microscopic nature, i.e., the phonon-assisted scattering, the modified rates should also have different energy dependence, which may strongly enhance their contribution in the Hall photoresponce in situations when the linear terms $\gamma_3^{(2)}$ are strongly suppressed, for instance, via cancellations discussed below Eq.~(\ref{durnev16}). The heating effects are directly related to the radiation absorption that scales as $(\omega\tau_1)^{-2}$. Thus, not only high gate voltages, but also a reduction of frequency should make the nonlinear contributions more prominent. This is indeed detected in experiments, demonstrating that at low frequencies the superlinear $I^2$ photoresponse becomes dominating at substantially lower intensities, see Fig.~\ref{frequency2}.

\section{Summary}

Our experiments demonstrate that excitation of graphene by circularly polarized terahertz radiation results in a helicity sensitive transverse photoconductivity originating from the scattering-assisted intraband absorption. Depending on the gate voltage, the transverse photoconductivity signal exhibits either linear or quadratic growth with the radiation intensity $I$. In the former case our results are well captured by the recently developed analytical theory~\cite{Durnev2021a} taking into account the alignment of electron momenta by combined action of THz and static electric fields as well as the dynamic heating and cooling of the electron gas. In particular, this theory is capable to explain a nonmonotonic dependence of the circular photoconductivity on the gate voltage at low intensities, with the position of the maximal signal reflecting the change in the microscopic nature of scattering at intermediate carrier densities. In our experiments, the linear-in-$I$ terms in the photoconductivity become strongly suppressed at high gate voltages. The photoconductivity signal here is dominated by contributions scaling quadratically with $I$, which consistently brings the position of the maximal signal to higher gate voltages at higher intensities. We discuss this unusual behavior in terms of an interplay between the second- and fourth-order effects in the radiation electric field, with emphasis on the heating effects that determine the longitudinal photoconductivity response in our experiments, and may also play an important role in formation of the high-intensity quadratic transverse photoresponse.

\section*{Acknowledgments}
We thank M.~V. Durnev and S.~A. Tarasenko for helpful discussions. 
The support from the FLAG-ERA program (project DeMeGRaS, project GA501/16-1 of the Deutsche Forschungsgemeinschaft, DFG), the Elite Network of Bavaria (K-NW-2013-247), and the Volkswagen Stiftung Program 97738 is gratefully acknowledged. 
ID acknowledges support of the Deutsche Forschungsgemeinschaft (DFG project DM1/5-1). 
SG acknowledges support of the IRAP Programme of the Foundation for Polish Science (grant MAB/2018/9, project CENTERA). 
K.W. and T.T. acknowledge support from the Elemental Strategy Initiative conducted by the MEXT, Japan (Grant Number JPMXP0112101001) and  JSPS KAKENHI (Grant Numbers 19H05790, 20H00354 and 21H05233).

\newpage
\FloatBarrier

\section{supplementary materials}

\subsection{Comparison of Hall photoconductivity and photocurrent response}	

\begin{figure}
	\centering
	\includegraphics[]{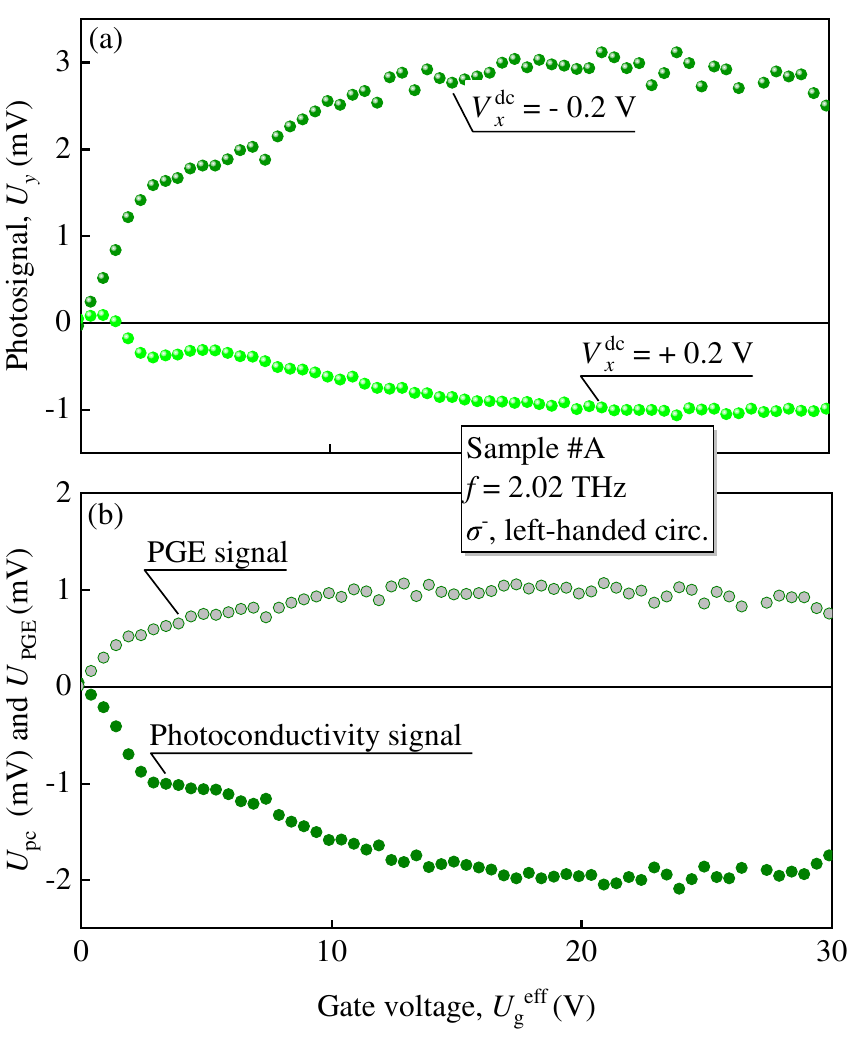}
	\caption{(a) Gate voltage dependence of the Hall photosignal $U_y$ recorded at bias voltages $V_x^\text{dc}=-0.2$~V and $V_x^\text{dc}=0.2$~V. (b) The odd and even parts of the data in panel (a) extracted as $U_{\rm pc}=[U_y(V_x^\text{dc})-U_y(-V_x^\text{dc})]/2$ and $U_{\rm PGE}=[U_y(V_x^\text{dc})+U_y(-V_x^\text{dc})]/2$ that yield the transverse photoconductivity and photogalvanic (PGE) signal, respectively. Measurements were performed on sample \#A under $f=2.02$\,THz left-handed circularly polarized radiation.
}
	\label{current1}
\end{figure}

Besides the change of the sample {\it dc} conductivity, excitation of graphene with terahertz radiation may also produce photogalvanic currents~\cite{Glazov2014,Candussio2021a} which can contribute to the total signal. 
For the measurements of the photoconductivity the sample was biased by a positive or negative dc bias voltage $V_x^\text{dc}$, see Figs.~\ref{current1}(a). In this kind of measurements, the photosignal $U_y$ consists of two contributions associated with the generation of photocurrent and the change of conductivity upon irradiation. The former one is independent of the bias polarity, whereas the latter should change its sign upon reversing $V_x^\text{dc}$. Accordingly, in Eq.~(\ref{eqodd}) we extracted the photoconductivity signal, $U_{\rm pc}$, as the odd part of the voltage signal with respect to the  bias voltage, $U_{\rm pc}= [U_y(V_x^\text{dc})-U_y(-V_x^\text{dc})]/2$. Correspondingly, the background even part of the signal, representing the photogalvanic effect (PGE), is given by
\begin{equation}
U_{\rm PGE}= \frac{U_y(V_x^\text{dc})+U_y(-V_x^\text{dc}) }{2}.
\end{equation}
Figure~\ref{current1}(b) shows the corresponding even and odd parts of the Hall photosignal as a function of the gate voltage. It demonstrates that in this example for $f=$2.2~THz radiation the transverse photoconductivity signal is about two times larger than that of the PGE. Note that in the present paper we focus on the photoconductivity and thus do not further discuss the data representing the photogalvanic currents. Experimental results and mechanisms of the PGE were presented in Ref.~\cite{Candussio2021a} where  the THz radiation-induced PGE in similar graphene samples was studied.

\subsection{Longitudinal photoconductivity}	

\begin{figure}
	\centering
	\includegraphics[]{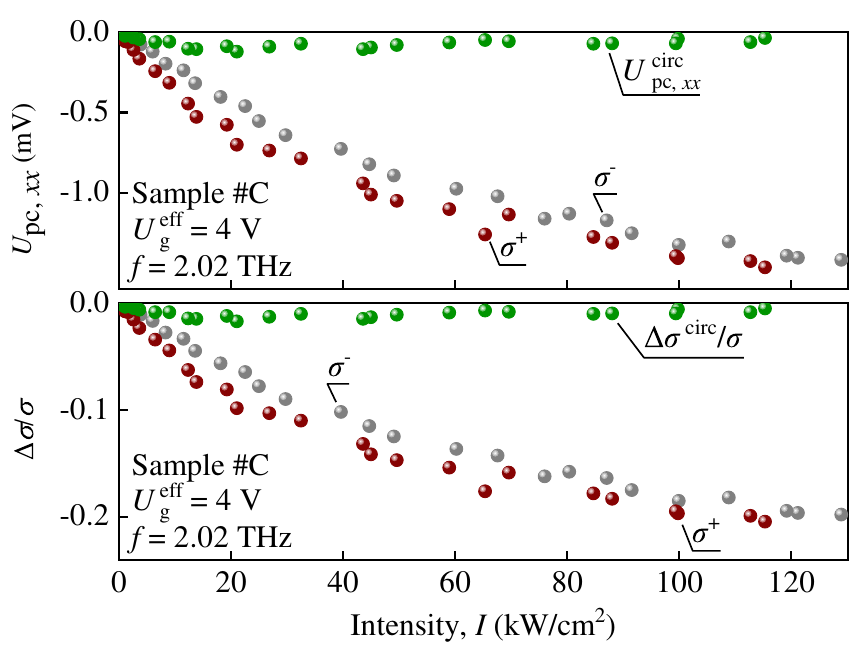}
	\caption{ The intensity dependence of (a) the longitudinal two-terminal photoconductivity signal $U_{{\rm pc},\,xx}$ and (b) the corresponding normalized longitudinal photoconductivity $\Delta\sigma/\sigma$ obtained for both helicities of $f=2.02$\,THz radiation. The difference of the photosignals for two helicities divided by two is also shown in both panels (marked by superscript circ) demonstrating that the helicity-dependent part of the longitudinal photoconductivity is vanishingly small in comparison with the total signal. These data were obtained on sample \#C, with the gate voltage fixed at $\Ug = 4$\,V, and the bias voltage $V_x^\text{dc}=\pm 0.3$~V.
	}
	\label{intdep_oldsample}  
\end{figure}

Figure~\ref{intdep_oldsample}(a) shows the longitudinal photoconductivity detected in two-terminal measurements using two contacts along the Hall bar in sample \#C, see Fig.~\ref{transport}~(a) and (b). Here the longitudinal photoconductivity signal $U_{{\rm pc},\,xx}=[U_x(V_x^\text{dc})-U_x(-V_x^\text{dc})]/2$ is shown, calculated using the voltage drops $U_x$ measured at $V_x^{\rm dc}=\pm 0.3$~V for two circular polarizations. Measuring the intensity dependence of such response we detected that at low intensities it grows linearly with $I$ and saturates at high intensities. Figure~\ref{intdep_oldsample}(b) shows the calculated relative photo-induced change of the longitudinal conductivity $\Delta\sigma/\sigma$ in units of the dark conductivity $\sigma$. The conductivity decreases upon irradiation. This behaviour is consistent with the  negative $\mu-$photoconductivity mechanism  which implies that the heating of charge carriers reduces their mobility, see, e.g., Ref.~\cite{Ganichev2005}. The observed decrease of the carrier mobility with increasing electron gas temperature is in agreement with the transport measurements (see, e.g., Ref.~\cite{Sarkar2015}) and originates from the scattering on acoustic phonons \cite{Wang2013}. 
	
For the terahertz radiation and in the used range of radiation intensities, the saturation of the photoconductivity response in gated samples is caused by the absorption bleaching~\cite{Ganichev2002,Candussio2021a,Danilov2021}. The bleaching of the Drude-like radiation absorption in monolayer graphene has been recently studied by means of the nonlinear ultrafast  \cite{Mics2015} and photogalvanic \cite{Candussio2021a} THz spectroscopy. The ranges of radiation frequencies (0.4–1.2~THz) and electric fields  (2 – 100~kV/cm) used in these studies are similar to those in our work. It has been shown that the absorption bleaching is caused by electron gas heating followed by the energy relaxation and is well described by an empirical formula 
\begin{equation} \label{eq2}
U_{{\rm pc},\,xx} \propto \Delta\sigma \propto I/(I + I_s)\, ,
\end{equation}
where $I_s$ is the saturation intensity \cite{Candussio2021a}. Comparison of the longitudinal photoresponses to the right- and left-handed circularly polarized radiation in Fig.~\ref{intdep_oldsample} shows that the signal does not depend on the radiation helicity. Indeed, the helicity-dependent part $U_{{\rm pc},\,xx}^{\rm circ}$ [defined analogous to Eq.~(\ref{eqcirc}) as one half of the difference between signals for opposite helicities] is close to zero, in sharp contrast to the transverse photoconductivity, for which the signal changes sign for the opposite radiation helicity. A small difference between the signals for opposite helicities, still visible in Fig.~\ref{intdep_oldsample}, is most probably caused by imperfections of the implemented $\lambda/4$-plate.

\bibliography{all_bib}

\begin{thebibliography}{51}%
\makeatletter
\providecommand \@ifxundefined [1]{%
 \@ifx{#1\undefined}
}%
\providecommand \@ifnum [1]{%
 \ifnum #1\expandafter \@firstoftwo
 \else \expandafter \@secondoftwo
 \fi
}%
\providecommand \@ifx [1]{%
 \ifx #1\expandafter \@firstoftwo
 \else \expandafter \@secondoftwo
 \fi
}%
\providecommand \natexlab [1]{#1}%
\providecommand \enquote  [1]{``#1''}%
\providecommand \bibnamefont  [1]{#1}%
\providecommand \bibfnamefont [1]{#1}%
\providecommand \citenamefont [1]{#1}%
\providecommand \href@noop [0]{\@secondoftwo}%
\providecommand \href [0]{\begingroup \@sanitize@url \@href}%
\providecommand \@href[1]{\@@startlink{#1}\@@href}%
\providecommand \@@href[1]{\endgroup#1\@@endlink}%
\providecommand \@sanitize@url [0]{\catcode `\\12\catcode `\$12\catcode
  `\&12\catcode `\#12\catcode `\^12\catcode `\_12\catcode `\%12\relax}%
\providecommand \@@startlink[1]{}%
\providecommand \@@endlink[0]{}%
\providecommand \url  [0]{\begingroup\@sanitize@url \@url }%
\providecommand \@url [1]{\endgroup\@href {#1}{\urlprefix }}%
\providecommand \urlprefix  [0]{URL }%
\providecommand \Eprint [0]{\href }%
\providecommand \doibase [0]{https://doi.org/}%
\providecommand \selectlanguage [0]{\@gobble}%
\providecommand \bibinfo  [0]{\@secondoftwo}%
\providecommand \bibfield  [0]{\@secondoftwo}%
\providecommand \translation [1]{[#1]}%
\providecommand \BibitemOpen [0]{}%
\providecommand \bibitemStop [0]{}%
\providecommand \bibitemNoStop [0]{.\EOS\space}%
\providecommand \EOS [0]{\spacefactor3000\relax}%
\providecommand \BibitemShut  [1]{\csname bibitem#1\endcsname}%
\let\auto@bib@innerbib\@empty
\bibitem [{\citenamefont {Xia}\ \emph {et~al.}(2009)\citenamefont {Xia},
  \citenamefont {Mueller}, \citenamefont {ming Lin}, \citenamefont
  {Valdes-Garcia},\ and\ \citenamefont {Avouris}}]{Xia2009a}%
  \BibitemOpen
  \bibfield  {author} {\bibinfo {author} {\bibfnamefont {F.}~\bibnamefont
  {Xia}}, \bibinfo {author} {\bibfnamefont {T.}~\bibnamefont {Mueller}},
  \bibinfo {author} {\bibfnamefont {Y.}~\bibnamefont {ming Lin}}, \bibinfo
  {author} {\bibfnamefont {A.}~\bibnamefont {Valdes-Garcia}},\ and\ \bibinfo
  {author} {\bibfnamefont {P.}~\bibnamefont {Avouris}},\ }\bibfield  {title}
  {\bibinfo {title} {Ultrafast graphene photodetector},\ }\href
  {https://doi.org/10.1038/nnano.2009.292} {\bibfield  {journal} {\bibinfo
  {journal} {Nat. Nanotechnol.}\ }\textbf {\bibinfo {volume} {4}},\ \bibinfo
  {pages} {839} (\bibinfo {year} {2009})}\BibitemShut {NoStop}%
\bibitem [{\citenamefont {Bonaccorso}\ \emph {et~al.}(2010)\citenamefont
  {Bonaccorso}, \citenamefont {Sun}, \citenamefont {Hasan},\ and\ \citenamefont
  {Ferrari}}]{Bonaccorso2010}%
  \BibitemOpen
  \bibfield  {author} {\bibinfo {author} {\bibfnamefont {F.}~\bibnamefont
  {Bonaccorso}}, \bibinfo {author} {\bibfnamefont {Z.}~\bibnamefont {Sun}},
  \bibinfo {author} {\bibfnamefont {T.}~\bibnamefont {Hasan}},\ and\ \bibinfo
  {author} {\bibfnamefont {A.~C.}\ \bibnamefont {Ferrari}},\ }\bibfield
  {title} {\bibinfo {title} {Graphene photonics and optoelectronics},\ }\href
  {https://doi.org/10.1038/nphoton.2010.186} {\bibfield  {journal} {\bibinfo
  {journal} {Nat. Photonics}\ }\textbf {\bibinfo {volume} {4}},\ \bibinfo
  {pages} {611} (\bibinfo {year} {2010})}\BibitemShut {NoStop}%
\bibitem [{\citenamefont {Liu}\ \emph {et~al.}(2011)\citenamefont {Liu},
  \citenamefont {Yin}, \citenamefont {Ulin-Avila}, \citenamefont {Geng},
  \citenamefont {Zentgraf}, \citenamefont {Ju}, \citenamefont {Wang},\ and\
  \citenamefont {Zhang}}]{Liu2011}%
  \BibitemOpen
  \bibfield  {author} {\bibinfo {author} {\bibfnamefont {M.}~\bibnamefont
  {Liu}}, \bibinfo {author} {\bibfnamefont {X.}~\bibnamefont {Yin}}, \bibinfo
  {author} {\bibfnamefont {E.}~\bibnamefont {Ulin-Avila}}, \bibinfo {author}
  {\bibfnamefont {B.}~\bibnamefont {Geng}}, \bibinfo {author} {\bibfnamefont
  {T.}~\bibnamefont {Zentgraf}}, \bibinfo {author} {\bibfnamefont
  {L.}~\bibnamefont {Ju}}, \bibinfo {author} {\bibfnamefont {F.}~\bibnamefont
  {Wang}},\ and\ \bibinfo {author} {\bibfnamefont {X.}~\bibnamefont {Zhang}},\
  }\bibfield  {title} {\bibinfo {title} {A graphene-based broadband optical
  modulator},\ }\href {https://doi.org/10.1038/nature10067} {\bibfield
  {journal} {\bibinfo  {journal} {Nature}\ }\textbf {\bibinfo {volume} {474}},\
  \bibinfo {pages} {64} (\bibinfo {year} {2011})}\BibitemShut {NoStop}%
\bibitem [{\citenamefont {Echtermeyer}\ \emph {et~al.}(2011)\citenamefont
  {Echtermeyer}, \citenamefont {Britnell}, \citenamefont {Jasnos},
  \citenamefont {Lombardo}, \citenamefont {Gorbachev}, \citenamefont
  {Grigorenko}, \citenamefont {Geim}, \citenamefont {Ferrari},\ and\
  \citenamefont {Novoselov}}]{Echtermeyer2011}%
  \BibitemOpen
  \bibfield  {author} {\bibinfo {author} {\bibfnamefont {T.~J.}\ \bibnamefont
  {Echtermeyer}}, \bibinfo {author} {\bibfnamefont {L.}~\bibnamefont
  {Britnell}}, \bibinfo {author} {\bibfnamefont {P.~K.}\ \bibnamefont
  {Jasnos}}, \bibinfo {author} {\bibfnamefont {A.}~\bibnamefont {Lombardo}},
  \bibinfo {author} {\bibfnamefont {R.~V.}\ \bibnamefont {Gorbachev}}, \bibinfo
  {author} {\bibfnamefont {A.~N.}\ \bibnamefont {Grigorenko}}, \bibinfo
  {author} {\bibfnamefont {A.~K.}\ \bibnamefont {Geim}}, \bibinfo {author}
  {\bibfnamefont {A.~C.}\ \bibnamefont {Ferrari}},\ and\ \bibinfo {author}
  {\bibfnamefont {K.~S.}\ \bibnamefont {Novoselov}},\ }\bibfield  {title}
  {\bibinfo {title} {Strong plasmonic enhancement of photovoltage in
  graphene},\ }\href {https://doi.org/10.1038/ncomms1464} {\bibfield  {journal}
  {\bibinfo  {journal} {Nat. Commun.}\ }\textbf {\bibinfo {volume} {2}},\
  \bibinfo {pages} {458} (\bibinfo {year} {2011})}\BibitemShut {NoStop}%
\bibitem [{\citenamefont {Koppens}\ \emph {et~al.}(2011)\citenamefont
  {Koppens}, \citenamefont {Chang},\ and\ \citenamefont
  {de~Abajo}}]{Koppens2011}%
  \BibitemOpen
  \bibfield  {author} {\bibinfo {author} {\bibfnamefont {F.~H.~L.}\
  \bibnamefont {Koppens}}, \bibinfo {author} {\bibfnamefont {D.~E.}\
  \bibnamefont {Chang}},\ and\ \bibinfo {author} {\bibfnamefont {F.~J.~G.}\
  \bibnamefont {de~Abajo}},\ }\bibfield  {title} {\bibinfo {title} {Graphene
  plasmonics: A platform for strong light{\textendash}matter interactions},\
  }\href {https://doi.org/10.1021/nl201771h} {\bibfield  {journal} {\bibinfo
  {journal} {Nano Lett.}\ }\textbf {\bibinfo {volume} {11}},\ \bibinfo {pages}
  {3370} (\bibinfo {year} {2011})}\BibitemShut {NoStop}%
\bibitem [{\citenamefont {Vicarelli}\ \emph {et~al.}(2012)\citenamefont
  {Vicarelli}, \citenamefont {Vitiello}, \citenamefont {Coquillat},
  \citenamefont {Lombardo}, \citenamefont {Ferrari}, \citenamefont {Knap},
  \citenamefont {Polini}, \citenamefont {Pellegrini},\ and\ \citenamefont
  {Tredicucci}}]{Vicarelli2012}%
  \BibitemOpen
  \bibfield  {author} {\bibinfo {author} {\bibfnamefont {L.}~\bibnamefont
  {Vicarelli}}, \bibinfo {author} {\bibfnamefont {M.~S.}\ \bibnamefont
  {Vitiello}}, \bibinfo {author} {\bibfnamefont {D.}~\bibnamefont {Coquillat}},
  \bibinfo {author} {\bibfnamefont {A.}~\bibnamefont {Lombardo}}, \bibinfo
  {author} {\bibfnamefont {A.~C.}\ \bibnamefont {Ferrari}}, \bibinfo {author}
  {\bibfnamefont {W.}~\bibnamefont {Knap}}, \bibinfo {author} {\bibfnamefont
  {M.}~\bibnamefont {Polini}}, \bibinfo {author} {\bibfnamefont
  {V.}~\bibnamefont {Pellegrini}},\ and\ \bibinfo {author} {\bibfnamefont
  {A.}~\bibnamefont {Tredicucci}},\ }\bibfield  {title} {\bibinfo {title}
  {Graphene field-effect transistors as room-temperature terahertz detectors},\
  }\href {https://doi.org/10.1038/nmat3417} {\bibfield  {journal} {\bibinfo
  {journal} {Nat. Mater.}\ }\textbf {\bibinfo {volume} {11}},\ \bibinfo {pages}
  {865} (\bibinfo {year} {2012})}\BibitemShut {NoStop}%
\bibitem [{\citenamefont {Engel}\ \emph {et~al.}(2012)\citenamefont {Engel},
  \citenamefont {Steiner}, \citenamefont {Lombardo}, \citenamefont {Ferrari},
  \citenamefont {v.~Löhneysen}, \citenamefont {Avouris},\ and\ \citenamefont
  {Krupke}}]{Engel2012}%
  \BibitemOpen
  \bibfield  {author} {\bibinfo {author} {\bibfnamefont {M.}~\bibnamefont
  {Engel}}, \bibinfo {author} {\bibfnamefont {M.}~\bibnamefont {Steiner}},
  \bibinfo {author} {\bibfnamefont {A.}~\bibnamefont {Lombardo}}, \bibinfo
  {author} {\bibfnamefont {A.~C.}\ \bibnamefont {Ferrari}}, \bibinfo {author}
  {\bibfnamefont {H.}~\bibnamefont {v.~Löhneysen}}, \bibinfo {author}
  {\bibfnamefont {P.}~\bibnamefont {Avouris}},\ and\ \bibinfo {author}
  {\bibfnamefont {R.}~\bibnamefont {Krupke}},\ }\bibfield  {title} {\bibinfo
  {title} {Light{\textendash}matter interaction in a microcavity-controlled
  graphene transistor},\ }\bibfield  {journal} {\bibinfo  {journal} {Nat.
  Commun.}\ }\textbf {\bibinfo {volume} {3}},\ \href
  {https://doi.org/10.1038/ncomms1911} {10.1038/ncomms1911} (\bibinfo {year}
  {2012})\BibitemShut {NoStop}%
\bibitem [{\citenamefont {Grigorenko}\ \emph {et~al.}(2012)\citenamefont
  {Grigorenko}, \citenamefont {Polini},\ and\ \citenamefont
  {Novoselov}}]{Grigorenko2012}%
  \BibitemOpen
  \bibfield  {author} {\bibinfo {author} {\bibfnamefont {A.~N.}\ \bibnamefont
  {Grigorenko}}, \bibinfo {author} {\bibfnamefont {M.}~\bibnamefont {Polini}},\
  and\ \bibinfo {author} {\bibfnamefont {K.~S.}\ \bibnamefont {Novoselov}},\
  }\bibfield  {title} {\bibinfo {title} {Graphene plasmonics},\ }\href
  {https://doi.org/10.1038/nphoton.2012.262} {\bibfield  {journal} {\bibinfo
  {journal} {Nat. Photonics}\ }\textbf {\bibinfo {volume} {6}},\ \bibinfo
  {pages} {749} (\bibinfo {year} {2012})}\BibitemShut {NoStop}%
\bibitem [{\citenamefont {Bao}\ and\ \citenamefont {Loh}(2012)}]{Bao2012}%
  \BibitemOpen
  \bibfield  {author} {\bibinfo {author} {\bibfnamefont {Q.}~\bibnamefont
  {Bao}}\ and\ \bibinfo {author} {\bibfnamefont {K.~P.}\ \bibnamefont {Loh}},\
  }\bibfield  {title} {\bibinfo {title} {Graphene photonics, plasmonics, and
  broadband optoelectronic devices},\ }\href
  {https://doi.org/10.1021/nn300989g} {\bibfield  {journal} {\bibinfo
  {journal} {{ACS} Nano}\ }\textbf {\bibinfo {volume} {6}},\ \bibinfo {pages}
  {3677} (\bibinfo {year} {2012})}\BibitemShut {NoStop}%
\bibitem [{\citenamefont {Jariwala}\ \emph {et~al.}(2013)\citenamefont
  {Jariwala}, \citenamefont {Sangwan}, \citenamefont {Lauhon}, \citenamefont
  {Marks},\ and\ \citenamefont {Hersam}}]{Jariwala2013}%
  \BibitemOpen
  \bibfield  {author} {\bibinfo {author} {\bibfnamefont {D.}~\bibnamefont
  {Jariwala}}, \bibinfo {author} {\bibfnamefont {V.~K.}\ \bibnamefont
  {Sangwan}}, \bibinfo {author} {\bibfnamefont {L.~J.}\ \bibnamefont {Lauhon}},
  \bibinfo {author} {\bibfnamefont {T.~J.}\ \bibnamefont {Marks}},\ and\
  \bibinfo {author} {\bibfnamefont {M.~C.}\ \bibnamefont {Hersam}},\ }\bibfield
   {title} {\bibinfo {title} {Carbon nanomaterials for electronics,
  optoelectronics, photovoltaics, and sensing},\ }\href
  {https://doi.org/10.1039/c2cs35335k} {\bibfield  {journal} {\bibinfo
  {journal} {Chem. Soc. Rev.}\ }\textbf {\bibinfo {volume} {42}},\ \bibinfo
  {pages} {2824} (\bibinfo {year} {2013})}\BibitemShut {NoStop}%
\bibitem [{\citenamefont {Glazov}\ and\ \citenamefont
  {Ganichev}(2014)}]{Glazov2014}%
  \BibitemOpen
  \bibfield  {author} {\bibinfo {author} {\bibfnamefont {M.~M.}\ \bibnamefont
  {Glazov}}\ and\ \bibinfo {author} {\bibfnamefont {S.~D.}\ \bibnamefont
  {Ganichev}},\ }\bibfield  {title} {\bibinfo {title} {High frequency electric
  field induced nonlinear effects in graphene},\ }\href
  {https://doi.org/10.1016/j.physrep.2013.10.003} {\bibfield  {journal}
  {\bibinfo  {journal} {Phys. Rep.}\ }\textbf {\bibinfo {volume} {535}},\
  \bibinfo {pages} {101} (\bibinfo {year} {2014})}\BibitemShut {NoStop}%
\bibitem [{\citenamefont {Koppens}\ \emph {et~al.}(2014)\citenamefont
  {Koppens}, \citenamefont {Mueller}, \citenamefont {Avouris}, \citenamefont
  {Ferrari}, \citenamefont {Vitiello},\ and\ \citenamefont
  {Polini}}]{Koppens2014}%
  \BibitemOpen
  \bibfield  {author} {\bibinfo {author} {\bibfnamefont {F.~H.~L.}\
  \bibnamefont {Koppens}}, \bibinfo {author} {\bibfnamefont {T.}~\bibnamefont
  {Mueller}}, \bibinfo {author} {\bibfnamefont {P.}~\bibnamefont {Avouris}},
  \bibinfo {author} {\bibfnamefont {A.~C.}\ \bibnamefont {Ferrari}}, \bibinfo
  {author} {\bibfnamefont {M.~S.}\ \bibnamefont {Vitiello}},\ and\ \bibinfo
  {author} {\bibfnamefont {M.}~\bibnamefont {Polini}},\ }\bibfield  {title}
  {\bibinfo {title} {Photodetectors based on graphene, other two-dimensional
  materials and hybrid systems},\ }\href
  {http://dx.doi.org/10.1038/nnano.2014.215} {\bibfield  {journal} {\bibinfo
  {journal} {Nat. Nanotechnol.}\ }\textbf {\bibinfo {volume} {9}},\ \bibinfo
  {pages} {780} (\bibinfo {year} {2014})}\BibitemShut {NoStop}%
\bibitem [{\citenamefont {Sun}\ and\ \citenamefont {Chang}(2014)}]{Sun2014}%
  \BibitemOpen
  \bibfield  {author} {\bibinfo {author} {\bibfnamefont {Z.}~\bibnamefont
  {Sun}}\ and\ \bibinfo {author} {\bibfnamefont {H.}~\bibnamefont {Chang}},\
  }\bibfield  {title} {\bibinfo {title} {Graphene and graphene-like
  two-dimensional materials in photodetection: Mechanisms and methodology},\
  }\href {https://doi.org/10.1021/nn500508c} {\bibfield  {journal} {\bibinfo
  {journal} {{ACS} Nano}\ }\textbf {\bibinfo {volume} {8}},\ \bibinfo {pages}
  {4133} (\bibinfo {year} {2014})}\BibitemShut {NoStop}%
\bibitem [{\citenamefont {Mueller}\ \emph {et~al.}(2014)\citenamefont
  {Mueller}, \citenamefont {Ferrari}, \citenamefont {Koppens}, \citenamefont
  {Xia},\ and\ \citenamefont {Xu}}]{Mueller2014}%
  \BibitemOpen
  \bibfield  {author} {\bibinfo {author} {\bibfnamefont {T.}~\bibnamefont
  {Mueller}}, \bibinfo {author} {\bibfnamefont {A.~C.}\ \bibnamefont
  {Ferrari}}, \bibinfo {author} {\bibfnamefont {F.}~\bibnamefont {Koppens}},
  \bibinfo {author} {\bibfnamefont {F.}~\bibnamefont {Xia}},\ and\ \bibinfo
  {author} {\bibfnamefont {X.}~\bibnamefont {Xu}},\ }\bibfield  {title}
  {\bibinfo {title} {Introduction to the issue on graphene optoelectronics},\
  }\href {https://doi.org/10.1109/jstqe.2013.2297054} {\bibfield  {journal}
  {\bibinfo  {journal} {{IEEE} Journal of Selected Topics in Quantum
  Electronics}\ }\textbf {\bibinfo {volume} {20}},\ \bibinfo {pages} {6}
  (\bibinfo {year} {2014})}\BibitemShut {NoStop}%
\bibitem [{\citenamefont {Sun}\ \emph {et~al.}(2016)\citenamefont {Sun},
  \citenamefont {Martinez},\ and\ \citenamefont {Wang}}]{Sun2016}%
  \BibitemOpen
  \bibfield  {author} {\bibinfo {author} {\bibfnamefont {Z.}~\bibnamefont
  {Sun}}, \bibinfo {author} {\bibfnamefont {A.}~\bibnamefont {Martinez}},\ and\
  \bibinfo {author} {\bibfnamefont {F.}~\bibnamefont {Wang}},\ }\bibfield
  {title} {\bibinfo {title} {Optical modulators with 2d layered materials},\
  }\href {https://doi.org/10.1038/nphoton.2016.15} {\bibfield  {journal}
  {\bibinfo  {journal} {Nat. Photonics}\ }\textbf {\bibinfo {volume} {10}},\
  \bibinfo {pages} {227} (\bibinfo {year} {2016})}\BibitemShut {NoStop}%
\bibitem [{\citenamefont {Sanctis}\ \emph {et~al.}(2018)\citenamefont
  {Sanctis}, \citenamefont {Mehew}, \citenamefont {Craciun},\ and\
  \citenamefont {Russo}}]{Sanctis2018}%
  \BibitemOpen
  \bibfield  {author} {\bibinfo {author} {\bibfnamefont {A.~D.}\ \bibnamefont
  {Sanctis}}, \bibinfo {author} {\bibfnamefont {J.}~\bibnamefont {Mehew}},
  \bibinfo {author} {\bibfnamefont {M.}~\bibnamefont {Craciun}},\ and\ \bibinfo
  {author} {\bibfnamefont {S.}~\bibnamefont {Russo}},\ }\bibfield  {title}
  {\bibinfo {title} {Graphene-based light sensing: Fabrication,
  characterisation, physical properties and performance},\ }\href
  {https://doi.org/10.3390/ma11091762} {\bibfield  {journal} {\bibinfo
  {journal} {Materials}\ }\textbf {\bibinfo {volume} {11}},\ \bibinfo {pages}
  {1762} (\bibinfo {year} {2018})}\BibitemShut {NoStop}%
\bibitem [{\citenamefont {Wang}\ \emph {et~al.}(2019)\citenamefont {Wang},
  \citenamefont {Wu},\ and\ \citenamefont {Zhao}}]{Wang2019}%
  \BibitemOpen
  \bibfield  {author} {\bibinfo {author} {\bibfnamefont {Y.}~\bibnamefont
  {Wang}}, \bibinfo {author} {\bibfnamefont {W.}~\bibnamefont {Wu}},\ and\
  \bibinfo {author} {\bibfnamefont {Z.}~\bibnamefont {Zhao}},\ }\bibfield
  {title} {\bibinfo {title} {Recent progress and remaining challenges of 2d
  material-based terahertz detectors},\ }\href
  {https://doi.org/10.1016/j.infrared.2019.103024} {\bibfield  {journal}
  {\bibinfo  {journal} {Infrared Phys. Technol.}\ }\textbf {\bibinfo {volume}
  {102}},\ \bibinfo {pages} {103024} (\bibinfo {year} {2019})}\BibitemShut
  {NoStop}%
\bibitem [{\citenamefont {Tan}\ \emph {et~al.}(2020)\citenamefont {Tan},
  \citenamefont {Jiang}, \citenamefont {Wang}, \citenamefont {Yao},\ and\
  \citenamefont {Zhang}}]{Tan2020}%
  \BibitemOpen
  \bibfield  {author} {\bibinfo {author} {\bibfnamefont {T.}~\bibnamefont
  {Tan}}, \bibinfo {author} {\bibfnamefont {X.}~\bibnamefont {Jiang}}, \bibinfo
  {author} {\bibfnamefont {C.}~\bibnamefont {Wang}}, \bibinfo {author}
  {\bibfnamefont {B.}~\bibnamefont {Yao}},\ and\ \bibinfo {author}
  {\bibfnamefont {H.}~\bibnamefont {Zhang}},\ }\bibfield  {title} {\bibinfo
  {title} {2d material optoelectronics for information functional device
  applications: Status and challenges},\ }\href
  {https://doi.org/10.1002/advs.202000058} {\bibfield  {journal} {\bibinfo
  {journal} {Adv. Sci.}\ }\textbf {\bibinfo {volume} {7}},\ \bibinfo {pages}
  {2000058} (\bibinfo {year} {2020})}\BibitemShut {NoStop}%
\bibitem [{\citenamefont {Oka}\ and\ \citenamefont {Aoki}(2009)}]{Oka2009}%
  \BibitemOpen
  \bibfield  {author} {\bibinfo {author} {\bibfnamefont {T.}~\bibnamefont
  {Oka}}\ and\ \bibinfo {author} {\bibfnamefont {H.}~\bibnamefont {Aoki}},\
  }\bibfield  {title} {\bibinfo {title} {Photovoltaic hall effect in
  graphene},\ }\href {https://doi.org/10.1103/physrevb.79.081406} {\bibfield
  {journal} {\bibinfo  {journal} {Phys. Rev. B}\ }\textbf {\bibinfo {volume}
  {79}},\ \bibinfo {pages} {081406 (R)} (\bibinfo {year} {2009})}\BibitemShut
  {NoStop}%
\bibitem [{\citenamefont {Karch}\ \emph {et~al.}(2010)\citenamefont {Karch},
  \citenamefont {Olbrich}, \citenamefont {Schmalzbauer}, \citenamefont {Zoth},
  \citenamefont {Brinsteiner}, \citenamefont {Fehrenbacher}, \citenamefont
  {Wurstbauer}, \citenamefont {Glazov}, \citenamefont {Tarasenko},
  \citenamefont {Ivchenko}, \citenamefont {Weiss}, \citenamefont {Eroms},
  \citenamefont {Yakimova}, \citenamefont {Lara-Avila}, \citenamefont
  {Kubatkin},\ and\ \citenamefont {Ganichev}}]{Karch2010}%
  \BibitemOpen
  \bibfield  {author} {\bibinfo {author} {\bibfnamefont {J.}~\bibnamefont
  {Karch}}, \bibinfo {author} {\bibfnamefont {P.}~\bibnamefont {Olbrich}},
  \bibinfo {author} {\bibfnamefont {M.}~\bibnamefont {Schmalzbauer}}, \bibinfo
  {author} {\bibfnamefont {C.}~\bibnamefont {Zoth}}, \bibinfo {author}
  {\bibfnamefont {C.}~\bibnamefont {Brinsteiner}}, \bibinfo {author}
  {\bibfnamefont {M.}~\bibnamefont {Fehrenbacher}}, \bibinfo {author}
  {\bibfnamefont {U.}~\bibnamefont {Wurstbauer}}, \bibinfo {author}
  {\bibfnamefont {M.~M.}\ \bibnamefont {Glazov}}, \bibinfo {author}
  {\bibfnamefont {S.~A.}\ \bibnamefont {Tarasenko}}, \bibinfo {author}
  {\bibfnamefont {E.~L.}\ \bibnamefont {Ivchenko}}, \bibinfo {author}
  {\bibfnamefont {D.}~\bibnamefont {Weiss}}, \bibinfo {author} {\bibfnamefont
  {J.}~\bibnamefont {Eroms}}, \bibinfo {author} {\bibfnamefont
  {R.}~\bibnamefont {Yakimova}}, \bibinfo {author} {\bibfnamefont
  {S.}~\bibnamefont {Lara-Avila}}, \bibinfo {author} {\bibfnamefont
  {S.}~\bibnamefont {Kubatkin}},\ and\ \bibinfo {author} {\bibfnamefont
  {S.~D.}\ \bibnamefont {Ganichev}},\ }\bibfield  {title} {\bibinfo {title}
  {Dynamic hall effect driven by circularly polarized light in a graphene
  layer},\ }\href {https://doi.org/10.1103/PhysRevLett.105.227402} {\bibfield
  {journal} {\bibinfo  {journal} {Phys. Rev. Lett.}\ }\textbf {\bibinfo
  {volume} {105}},\ \bibinfo {pages} {227402} (\bibinfo {year}
  {2010})}\BibitemShut {NoStop}%
\bibitem [{\citenamefont {Karch}\ \emph {et~al.}(2011)\citenamefont {Karch},
  \citenamefont {Drexler}, \citenamefont {Olbrich}, \citenamefont
  {Fehrenbacher}, \citenamefont {Hirmer}, \citenamefont {Glazov}, \citenamefont
  {Tarasenko}, \citenamefont {Ivchenko}, \citenamefont {Birkner}, \citenamefont
  {Eroms}, \citenamefont {Weiss}, \citenamefont {Yakimova}, \citenamefont
  {Lara-Avila}, \citenamefont {Kubatkin}, \citenamefont {Ostler}, \citenamefont
  {Seyller},\ and\ \citenamefont {Ganichev}}]{Karch2011}%
  \BibitemOpen
  \bibfield  {author} {\bibinfo {author} {\bibfnamefont {J.}~\bibnamefont
  {Karch}}, \bibinfo {author} {\bibfnamefont {C.}~\bibnamefont {Drexler}},
  \bibinfo {author} {\bibfnamefont {P.}~\bibnamefont {Olbrich}}, \bibinfo
  {author} {\bibfnamefont {M.}~\bibnamefont {Fehrenbacher}}, \bibinfo {author}
  {\bibfnamefont {M.}~\bibnamefont {Hirmer}}, \bibinfo {author} {\bibfnamefont
  {M.~M.}\ \bibnamefont {Glazov}}, \bibinfo {author} {\bibfnamefont {S.~A.}\
  \bibnamefont {Tarasenko}}, \bibinfo {author} {\bibfnamefont {E.~L.}\
  \bibnamefont {Ivchenko}}, \bibinfo {author} {\bibfnamefont {B.}~\bibnamefont
  {Birkner}}, \bibinfo {author} {\bibfnamefont {J.}~\bibnamefont {Eroms}},
  \bibinfo {author} {\bibfnamefont {D.}~\bibnamefont {Weiss}}, \bibinfo
  {author} {\bibfnamefont {R.}~\bibnamefont {Yakimova}}, \bibinfo {author}
  {\bibfnamefont {S.}~\bibnamefont {Lara-Avila}}, \bibinfo {author}
  {\bibfnamefont {S.}~\bibnamefont {Kubatkin}}, \bibinfo {author}
  {\bibfnamefont {M.}~\bibnamefont {Ostler}}, \bibinfo {author} {\bibfnamefont
  {T.}~\bibnamefont {Seyller}},\ and\ \bibinfo {author} {\bibfnamefont {S.~D.}\
  \bibnamefont {Ganichev}},\ }\bibfield  {title} {\bibinfo {title} {Terahertz
  radiation driven chiral edge currents in graphene},\ }\href
  {https://doi.org/10.1103/PhysRevLett.107.276601} {\bibfield  {journal}
  {\bibinfo  {journal} {Phys. Rev. Lett.}\ }\textbf {\bibinfo {volume} {107}},\
  \bibinfo {pages} {276601} (\bibinfo {year} {2011})}\BibitemShut {NoStop}%
\bibitem [{\citenamefont {Jiang}\ \emph {et~al.}(2011)\citenamefont {Jiang},
  \citenamefont {Shalygin}, \citenamefont {Panevin}, \citenamefont {Danilov},
  \citenamefont {Glazov}, \citenamefont {Yakimova}, \citenamefont {Lara-Avila},
  \citenamefont {Kubatkin},\ and\ \citenamefont {Ganichev}}]{Jiang2011}%
  \BibitemOpen
  \bibfield  {author} {\bibinfo {author} {\bibfnamefont {C.}~\bibnamefont
  {Jiang}}, \bibinfo {author} {\bibfnamefont {V.~A.}\ \bibnamefont {Shalygin}},
  \bibinfo {author} {\bibfnamefont {V.~Y.}\ \bibnamefont {Panevin}}, \bibinfo
  {author} {\bibfnamefont {S.~N.}\ \bibnamefont {Danilov}}, \bibinfo {author}
  {\bibfnamefont {M.~M.}\ \bibnamefont {Glazov}}, \bibinfo {author}
  {\bibfnamefont {R.}~\bibnamefont {Yakimova}}, \bibinfo {author}
  {\bibfnamefont {S.}~\bibnamefont {Lara-Avila}}, \bibinfo {author}
  {\bibfnamefont {S.}~\bibnamefont {Kubatkin}},\ and\ \bibinfo {author}
  {\bibfnamefont {S.~D.}\ \bibnamefont {Ganichev}},\ }\bibfield  {title}
  {\bibinfo {title} {Helicity-dependent photocurrents in graphene layers
  excited by midinfrared radiation of a co$_2$ laser},\ }\href
  {https://doi.org/10.1103/physrevb.84.125429} {\bibfield  {journal} {\bibinfo
  {journal} {Phys. Rev. B}\ }\textbf {\bibinfo {volume} {84}},\ \bibinfo
  {pages} {125429} (\bibinfo {year} {2011})}\BibitemShut {NoStop}%
\bibitem [{\citenamefont {Kitagawa}\ \emph {et~al.}(2011)\citenamefont
  {Kitagawa}, \citenamefont {Oka}, \citenamefont {Brataas}, \citenamefont
  {Fu},\ and\ \citenamefont {Demler}}]{Kitagawa2011}%
  \BibitemOpen
  \bibfield  {author} {\bibinfo {author} {\bibfnamefont {T.}~\bibnamefont
  {Kitagawa}}, \bibinfo {author} {\bibfnamefont {T.}~\bibnamefont {Oka}},
  \bibinfo {author} {\bibfnamefont {A.}~\bibnamefont {Brataas}}, \bibinfo
  {author} {\bibfnamefont {L.}~\bibnamefont {Fu}},\ and\ \bibinfo {author}
  {\bibfnamefont {E.}~\bibnamefont {Demler}},\ }\bibfield  {title} {\bibinfo
  {title} {Transport properties of nonequilibrium systems under the application
  of light: Photoinduced quantum hall insulators without landau levels},\
  }\href {https://doi.org/10.1103/physrevb.84.235108} {\bibfield  {journal}
  {\bibinfo  {journal} {Phys. Rev. B}\ }\textbf {\bibinfo {volume} {84}},\
  \bibinfo {pages} {235108} (\bibinfo {year} {2011})}\BibitemShut {NoStop}%
\bibitem [{\citenamefont {Ivchenko}(2012)}]{Ivchenko2012}%
  \BibitemOpen
  \bibfield  {author} {\bibinfo {author} {\bibfnamefont {E.~L.}\ \bibnamefont
  {Ivchenko}},\ }\bibfield  {title} {\bibinfo {title} {Photoinduced currents in
  graphene and carbon nanotubes},\ }\href
  {https://doi.org/10.1002/pssb.201200081} {\bibfield  {journal} {\bibinfo
  {journal} {Phys. Status Solidi B}\ }\textbf {\bibinfo {volume} {249}},\
  \bibinfo {pages} {2538} (\bibinfo {year} {2012})}\BibitemShut {NoStop}%
\bibitem [{\citenamefont {Qian}\ \emph {et~al.}(2018)\citenamefont {Qian},
  \citenamefont {Cao}, \citenamefont {Wang}, \citenamefont {Shen},
  \citenamefont {Soci}, \citenamefont {Eginligil},\ and\ \citenamefont
  {Yu}}]{Qian2018}%
  \BibitemOpen
  \bibfield  {author} {\bibinfo {author} {\bibfnamefont {X.}~\bibnamefont
  {Qian}}, \bibinfo {author} {\bibfnamefont {B.}~\bibnamefont {Cao}}, \bibinfo
  {author} {\bibfnamefont {Z.}~\bibnamefont {Wang}}, \bibinfo {author}
  {\bibfnamefont {X.}~\bibnamefont {Shen}}, \bibinfo {author} {\bibfnamefont
  {C.}~\bibnamefont {Soci}}, \bibinfo {author} {\bibfnamefont {M.}~\bibnamefont
  {Eginligil}},\ and\ \bibinfo {author} {\bibfnamefont {T.}~\bibnamefont
  {Yu}},\ }\bibfield  {title} {\bibinfo {title} {Carrier density and light
  helicity dependence of photocurrent in mono- and bilayer graphene},\ }\href
  {https://doi.org/10.1088/1361-6641/aae2f1} {\bibfield  {journal} {\bibinfo
  {journal} {Semicond. Sci. Technol.}\ }\textbf {\bibinfo {volume} {33}},\
  \bibinfo {pages} {114008} (\bibinfo {year} {2018})}\BibitemShut {NoStop}%
\bibitem [{\citenamefont {Zhu}\ \emph {et~al.}(2019)\citenamefont {Zhu},
  \citenamefont {Yao}, \citenamefont {Huang}, \citenamefont {He}, \citenamefont
  {Quan}, \citenamefont {Li}, \citenamefont {Gu}, \citenamefont {Xu},\ and\
  \citenamefont {Ren}}]{Zhu2019}%
  \BibitemOpen
  \bibfield  {author} {\bibinfo {author} {\bibfnamefont {L.}~\bibnamefont
  {Zhu}}, \bibinfo {author} {\bibfnamefont {Z.}~\bibnamefont {Yao}}, \bibinfo
  {author} {\bibfnamefont {Y.}~\bibnamefont {Huang}}, \bibinfo {author}
  {\bibfnamefont {C.}~\bibnamefont {He}}, \bibinfo {author} {\bibfnamefont
  {B.}~\bibnamefont {Quan}}, \bibinfo {author} {\bibfnamefont {J.}~\bibnamefont
  {Li}}, \bibinfo {author} {\bibfnamefont {C.}~\bibnamefont {Gu}}, \bibinfo
  {author} {\bibfnamefont {X.}~\bibnamefont {Xu}},\ and\ \bibinfo {author}
  {\bibfnamefont {Z.}~\bibnamefont {Ren}},\ }\bibfield  {title} {\bibinfo
  {title} {Circular-photon-drag-effect-induced elliptically polarized terahertz
  emission from vertically grown graphene},\ }\href
  {https://doi.org/10.1103/physrevapplied.12.044063} {\bibfield  {journal}
  {\bibinfo  {journal} {Phys. Rev. Appl}\ }\textbf {\bibinfo {volume} {12}},\
  \bibinfo {pages} {044063} (\bibinfo {year} {2019})}\BibitemShut {NoStop}%
\bibitem [{\citenamefont {McIver}\ \emph {et~al.}(2020)\citenamefont {McIver},
  \citenamefont {Schulte}, \citenamefont {Stein}, \citenamefont {Matsuyama},
  \citenamefont {Jotzu}, \citenamefont {Meier},\ and\ \citenamefont
  {Cavalleri}}]{McIver2020}%
  \BibitemOpen
  \bibfield  {author} {\bibinfo {author} {\bibfnamefont {J.~W.}\ \bibnamefont
  {McIver}}, \bibinfo {author} {\bibfnamefont {B.}~\bibnamefont {Schulte}},
  \bibinfo {author} {\bibfnamefont {F.-U.}\ \bibnamefont {Stein}}, \bibinfo
  {author} {\bibfnamefont {T.}~\bibnamefont {Matsuyama}}, \bibinfo {author}
  {\bibfnamefont {G.}~\bibnamefont {Jotzu}}, \bibinfo {author} {\bibfnamefont
  {G.}~\bibnamefont {Meier}},\ and\ \bibinfo {author} {\bibfnamefont
  {A.}~\bibnamefont {Cavalleri}},\ }\bibfield  {title} {\bibinfo {title}
  {Light-induced anomalous hall effect in graphene},\ }\href
  {https://doi.org/10.1038/s41567-019-0698-y} {\bibfield  {journal} {\bibinfo
  {journal} {Nat. Phys.}\ }\textbf {\bibinfo {volume} {16}},\ \bibinfo {pages}
  {38} (\bibinfo {year} {2020})}\BibitemShut {NoStop}%
\bibitem [{\citenamefont {Sato}\ \emph {et~al.}(2019)\citenamefont {Sato},
  \citenamefont {McIver}, \citenamefont {Nuske}, \citenamefont {Tang},
  \citenamefont {Jotzu}, \citenamefont {Schulte}, \citenamefont {Hübener},
  \citenamefont {Giovannini}, \citenamefont {Mathey}, \citenamefont {Sentef},
  \citenamefont {Cavalleri},\ and\ \citenamefont {Rubio}}]{Sato2019}%
  \BibitemOpen
  \bibfield  {author} {\bibinfo {author} {\bibfnamefont {S.~A.}\ \bibnamefont
  {Sato}}, \bibinfo {author} {\bibfnamefont {J.~W.}\ \bibnamefont {McIver}},
  \bibinfo {author} {\bibfnamefont {M.}~\bibnamefont {Nuske}}, \bibinfo
  {author} {\bibfnamefont {P.}~\bibnamefont {Tang}}, \bibinfo {author}
  {\bibfnamefont {G.}~\bibnamefont {Jotzu}}, \bibinfo {author} {\bibfnamefont
  {B.}~\bibnamefont {Schulte}}, \bibinfo {author} {\bibfnamefont
  {H.}~\bibnamefont {Hübener}}, \bibinfo {author} {\bibfnamefont {U.~D.}\
  \bibnamefont {Giovannini}}, \bibinfo {author} {\bibfnamefont
  {L.}~\bibnamefont {Mathey}}, \bibinfo {author} {\bibfnamefont {M.~A.}\
  \bibnamefont {Sentef}}, \bibinfo {author} {\bibfnamefont {A.}~\bibnamefont
  {Cavalleri}},\ and\ \bibinfo {author} {\bibfnamefont {A.}~\bibnamefont
  {Rubio}},\ }\bibfield  {title} {\bibinfo {title} {Microscopic theory for the
  light-induced anomalous hall effect in graphene},\ }\href
  {https://doi.org/10.1103/physrevb.99.214302} {\bibfield  {journal} {\bibinfo
  {journal} {Phys. Rev. B}\ }\textbf {\bibinfo {volume} {99}},\ \bibinfo
  {pages} {214302} (\bibinfo {year} {2019})}\BibitemShut {NoStop}%
\bibitem [{\citenamefont {Matyushkin}\ \emph {et~al.}(2020)\citenamefont
  {Matyushkin}, \citenamefont {Danilov}, \citenamefont {Moskotin},
  \citenamefont {Belosevich}, \citenamefont {Kaurova}, \citenamefont {Rybin},
  \citenamefont {Obraztsova}, \citenamefont {Fedorov}, \citenamefont
  {Gorbenko}, \citenamefont {Kachorovskii},\ and\ \citenamefont
  {Ganichev}}]{Matyushkin2020}%
  \BibitemOpen
  \bibfield  {author} {\bibinfo {author} {\bibfnamefont {Y.}~\bibnamefont
  {Matyushkin}}, \bibinfo {author} {\bibfnamefont {S.}~\bibnamefont {Danilov}},
  \bibinfo {author} {\bibfnamefont {M.}~\bibnamefont {Moskotin}}, \bibinfo
  {author} {\bibfnamefont {V.}~\bibnamefont {Belosevich}}, \bibinfo {author}
  {\bibfnamefont {N.}~\bibnamefont {Kaurova}}, \bibinfo {author} {\bibfnamefont
  {M.}~\bibnamefont {Rybin}}, \bibinfo {author} {\bibfnamefont {E.~D.}\
  \bibnamefont {Obraztsova}}, \bibinfo {author} {\bibfnamefont
  {G.}~\bibnamefont {Fedorov}}, \bibinfo {author} {\bibfnamefont
  {I.}~\bibnamefont {Gorbenko}}, \bibinfo {author} {\bibfnamefont
  {V.}~\bibnamefont {Kachorovskii}},\ and\ \bibinfo {author} {\bibfnamefont
  {S.}~\bibnamefont {Ganichev}},\ }\bibfield  {title} {\bibinfo {title}
  {Helicity-sensitive plasmonic terahertz interferometer},\ }\href
  {https://doi.org/10.1021/acs.nanolett.0c02692} {\bibfield  {journal}
  {\bibinfo  {journal} {Nano Lett.}\ }\textbf {\bibinfo {volume} {20}},\
  \bibinfo {pages} {7296} (\bibinfo {year} {2020})}\BibitemShut {NoStop}%
\bibitem [{\citenamefont {Otteneder}\ \emph {et~al.}(2020)\citenamefont
  {Otteneder}, \citenamefont {Hubmann}, \citenamefont {Lu}, \citenamefont
  {Kozlov}, \citenamefont {Golub}, \citenamefont {Watanabe}, \citenamefont
  {Taniguchi}, \citenamefont {Efetov},\ and\ \citenamefont
  {Ganichev}}]{Otteneder2020}%
  \BibitemOpen
  \bibfield  {author} {\bibinfo {author} {\bibfnamefont {M.}~\bibnamefont
  {Otteneder}}, \bibinfo {author} {\bibfnamefont {S.}~\bibnamefont {Hubmann}},
  \bibinfo {author} {\bibfnamefont {X.}~\bibnamefont {Lu}}, \bibinfo {author}
  {\bibfnamefont {D.~A.}\ \bibnamefont {Kozlov}}, \bibinfo {author}
  {\bibfnamefont {L.~E.}\ \bibnamefont {Golub}}, \bibinfo {author}
  {\bibfnamefont {K.}~\bibnamefont {Watanabe}}, \bibinfo {author}
  {\bibfnamefont {T.}~\bibnamefont {Taniguchi}}, \bibinfo {author}
  {\bibfnamefont {D.~K.}\ \bibnamefont {Efetov}},\ and\ \bibinfo {author}
  {\bibfnamefont {S.~D.}\ \bibnamefont {Ganichev}},\ }\bibfield  {title}
  {\bibinfo {title} {Terahertz photogalvanics in twisted bilayer graphene close
  to the second magic angle},\ }\href
  {https://doi.org/10.1021/acs.nanolett.0c02474} {\bibfield  {journal}
  {\bibinfo  {journal} {Nano Lett.}\ }\textbf {\bibinfo {volume} {20}},\
  \bibinfo {pages} {7152} (\bibinfo {year} {2020})}\BibitemShut {NoStop}%
\bibitem [{\citenamefont {Candussio}\ \emph
  {et~al.}(2021{\natexlab{a}})\citenamefont {Candussio}, \citenamefont
  {Durnev}, \citenamefont {Slizovskiy}, \citenamefont {Jötten}, \citenamefont
  {Keil}, \citenamefont {Bel'kov}, \citenamefont {Yin}, \citenamefont {Yang},
  \citenamefont {Son}, \citenamefont {Mishchenko}, \citenamefont {Fal'ko},\
  and\ \citenamefont {Ganichev}}]{Candussio2021}%
  \BibitemOpen
  \bibfield  {author} {\bibinfo {author} {\bibfnamefont {S.}~\bibnamefont
  {Candussio}}, \bibinfo {author} {\bibfnamefont {M.~V.}\ \bibnamefont
  {Durnev}}, \bibinfo {author} {\bibfnamefont {S.}~\bibnamefont {Slizovskiy}},
  \bibinfo {author} {\bibfnamefont {T.}~\bibnamefont {Jötten}}, \bibinfo
  {author} {\bibfnamefont {J.}~\bibnamefont {Keil}}, \bibinfo {author}
  {\bibfnamefont {V.~V.}\ \bibnamefont {Bel'kov}}, \bibinfo {author}
  {\bibfnamefont {J.}~\bibnamefont {Yin}}, \bibinfo {author} {\bibfnamefont
  {Y.}~\bibnamefont {Yang}}, \bibinfo {author} {\bibfnamefont {S.-K.}\
  \bibnamefont {Son}}, \bibinfo {author} {\bibfnamefont {A.}~\bibnamefont
  {Mishchenko}}, \bibinfo {author} {\bibfnamefont {V.}~\bibnamefont {Fal'ko}},\
  and\ \bibinfo {author} {\bibfnamefont {S.~D.}\ \bibnamefont {Ganichev}},\
  }\bibfield  {title} {\bibinfo {title} {Edge photocurrent in bilayer graphene
  due to inter-landau-level transitions},\ }\href
  {https://doi.org/10.1103/physrevb.103.125408} {\bibfield  {journal} {\bibinfo
   {journal} {Phys. Rev. B}\ }\textbf {\bibinfo {volume} {103}},\ \bibinfo
  {pages} {125408} (\bibinfo {year} {2021}{\natexlab{a}})}\BibitemShut
  {NoStop}%
\bibitem [{\citenamefont {Candussio}\ \emph
  {et~al.}(2021{\natexlab{b}})\citenamefont {Candussio}, \citenamefont {Golub},
  \citenamefont {Bernreuter}, \citenamefont {Jötten}, \citenamefont
  {Rockinger}, \citenamefont {Watanabe}, \citenamefont {Taniguchi},
  \citenamefont {Eroms}, \citenamefont {Weiss},\ and\ \citenamefont
  {Ganichev}}]{Candussio2021a}%
  \BibitemOpen
  \bibfield  {author} {\bibinfo {author} {\bibfnamefont {S.}~\bibnamefont
  {Candussio}}, \bibinfo {author} {\bibfnamefont {L.~E.}\ \bibnamefont
  {Golub}}, \bibinfo {author} {\bibfnamefont {S.}~\bibnamefont {Bernreuter}},
  \bibinfo {author} {\bibfnamefont {T.}~\bibnamefont {Jötten}}, \bibinfo
  {author} {\bibfnamefont {T.}~\bibnamefont {Rockinger}}, \bibinfo {author}
  {\bibfnamefont {K.}~\bibnamefont {Watanabe}}, \bibinfo {author}
  {\bibfnamefont {T.}~\bibnamefont {Taniguchi}}, \bibinfo {author}
  {\bibfnamefont {J.}~\bibnamefont {Eroms}}, \bibinfo {author} {\bibfnamefont
  {D.}~\bibnamefont {Weiss}},\ and\ \bibinfo {author} {\bibfnamefont {S.~D.}\
  \bibnamefont {Ganichev}},\ }\bibfield  {title} {\bibinfo {title} {Nonlinear
  intensity dependence of edge photocurrents in graphene induced by terahertz
  radiation},\ }\href {https://doi.org/10.1103/physrevb.104.155404} {\bibfield
  {journal} {\bibinfo  {journal} {Phys. Rev. B}\ }\textbf {\bibinfo {volume}
  {104}},\ \bibinfo {pages} {155404} (\bibinfo {year}
  {2021}{\natexlab{b}})}\BibitemShut {NoStop}%
\bibitem [{\citenamefont {Durnev}\ and\ \citenamefont
  {Tarasenko}(2021)}]{Durnev2021}%
  \BibitemOpen
  \bibfield  {author} {\bibinfo {author} {\bibfnamefont {M.~V.}\ \bibnamefont
  {Durnev}}\ and\ \bibinfo {author} {\bibfnamefont {S.~A.}\ \bibnamefont
  {Tarasenko}},\ }\bibfield  {title} {\bibinfo {title} {Edge photogalvanic
  effect caused by optical alignment of carrier momenta in two-dimensional
  dirac materials},\ }\href {https://doi.org/10.1103/physrevb.103.165411}
  {\bibfield  {journal} {\bibinfo  {journal} {Phys. Rev. B}\ }\textbf {\bibinfo
  {volume} {103}},\ \bibinfo {pages} {165411} (\bibinfo {year}
  {2021})}\BibitemShut {NoStop}%
\bibitem [{\citenamefont {Durnev}(2021)}]{Durnev2021a}%
  \BibitemOpen
  \bibfield  {author} {\bibinfo {author} {\bibfnamefont {M.~V.}\ \bibnamefont
  {Durnev}},\ }\bibfield  {title} {\bibinfo {title} {Photovoltaic hall effect
  in the two-dimensional electron gas: Kinetic theory},\ }\href
  {https://doi.org/10.1103/physrevb.104.085306} {\bibfield  {journal} {\bibinfo
   {journal} {Phys. Rev. B}\ }\textbf {\bibinfo {volume} {104}},\ \bibinfo
  {pages} {085306} (\bibinfo {year} {2021})}\BibitemShut {NoStop}%
\bibitem [{\citenamefont {Belinicher}\ and\ \citenamefont
  {Novikov}(1981)}]{Belinicher1981}%
  \BibitemOpen
  \bibfield  {author} {\bibinfo {author} {\bibfnamefont {V.~I.}\ \bibnamefont
  {Belinicher}}\ and\ \bibinfo {author} {\bibfnamefont {V.~N.}\ \bibnamefont
  {Novikov}},\ }\bibfield  {title} {\bibinfo {title} {Non-equilibrium
  photoconductivityand influence of external fields on the surface
  photogalvaniceffect},\ }\href@noop {} {\bibfield  {journal} {\bibinfo
  {journal} {Fiz.Tekh. Poluprovodn.}\ }\textbf {\bibinfo {volume} {15}},\
  \bibinfo {pages} {1957 [Sov. Phys. Semicond. 15, 1138 (1981)]} (\bibinfo
  {year} {1981})}\BibitemShut {NoStop}%
\bibitem [{\citenamefont {Dean}\ \emph {et~al.}(2010)\citenamefont {Dean},
  \citenamefont {Young}, \citenamefont {Meric}, \citenamefont {Lee},
  \citenamefont {Wang}, \citenamefont {Sorgenfrei}, \citenamefont {Watanabe},
  \citenamefont {Taniguchi}, \citenamefont {Kim}, \citenamefont {Shepard},\
  and\ \citenamefont {Hone}}]{Dean2010}%
  \BibitemOpen
  \bibfield  {author} {\bibinfo {author} {\bibfnamefont {C.~R.}\ \bibnamefont
  {Dean}}, \bibinfo {author} {\bibfnamefont {A.~F.}\ \bibnamefont {Young}},
  \bibinfo {author} {\bibfnamefont {I.}~\bibnamefont {Meric}}, \bibinfo
  {author} {\bibfnamefont {C.}~\bibnamefont {Lee}}, \bibinfo {author}
  {\bibfnamefont {L.}~\bibnamefont {Wang}}, \bibinfo {author} {\bibfnamefont
  {S.}~\bibnamefont {Sorgenfrei}}, \bibinfo {author} {\bibfnamefont
  {K.}~\bibnamefont {Watanabe}}, \bibinfo {author} {\bibfnamefont
  {T.}~\bibnamefont {Taniguchi}}, \bibinfo {author} {\bibfnamefont
  {P.}~\bibnamefont {Kim}}, \bibinfo {author} {\bibfnamefont {K.~L.}\
  \bibnamefont {Shepard}},\ and\ \bibinfo {author} {\bibfnamefont
  {J.}~\bibnamefont {Hone}},\ }\bibfield  {title} {\bibinfo {title} {Boron
  nitride substrates for high-quality graphene electronics},\ }\href
  {https://doi.org/10.1038/nnano.2010.172} {\bibfield  {journal} {\bibinfo
  {journal} {Nat. Nanotechnol.}\ }\textbf {\bibinfo {volume} {5}},\ \bibinfo
  {pages} {722} (\bibinfo {year} {2010})}\BibitemShut {NoStop}%
\bibitem [{\citenamefont {Wang}\ \emph {et~al.}(2013)\citenamefont {Wang},
  \citenamefont {Meric}, \citenamefont {Huang}, \citenamefont {Gao},
  \citenamefont {Gao}, \citenamefont {Tran}, \citenamefont {Taniguchi},
  \citenamefont {Watanabe}, \citenamefont {Campos}, \citenamefont {Muller},
  \citenamefont {Guo}, \citenamefont {Kim}, \citenamefont {Hone}, \citenamefont
  {Shepard},\ and\ \citenamefont {Dean}}]{Wang2013}%
  \BibitemOpen
  \bibfield  {author} {\bibinfo {author} {\bibfnamefont {L.}~\bibnamefont
  {Wang}}, \bibinfo {author} {\bibfnamefont {I.}~\bibnamefont {Meric}},
  \bibinfo {author} {\bibfnamefont {P.~Y.}\ \bibnamefont {Huang}}, \bibinfo
  {author} {\bibfnamefont {Q.}~\bibnamefont {Gao}}, \bibinfo {author}
  {\bibfnamefont {Y.}~\bibnamefont {Gao}}, \bibinfo {author} {\bibfnamefont
  {H.}~\bibnamefont {Tran}}, \bibinfo {author} {\bibfnamefont {T.}~\bibnamefont
  {Taniguchi}}, \bibinfo {author} {\bibfnamefont {K.}~\bibnamefont {Watanabe}},
  \bibinfo {author} {\bibfnamefont {L.~M.}\ \bibnamefont {Campos}}, \bibinfo
  {author} {\bibfnamefont {D.~A.}\ \bibnamefont {Muller}}, \bibinfo {author}
  {\bibfnamefont {J.}~\bibnamefont {Guo}}, \bibinfo {author} {\bibfnamefont
  {P.}~\bibnamefont {Kim}}, \bibinfo {author} {\bibfnamefont {J.}~\bibnamefont
  {Hone}}, \bibinfo {author} {\bibfnamefont {K.~L.}\ \bibnamefont {Shepard}},\
  and\ \bibinfo {author} {\bibfnamefont {C.~R.}\ \bibnamefont {Dean}},\
  }\bibfield  {title} {\bibinfo {title} {One-dimensional electrical contact to
  a two-dimensional material},\ }\href
  {https://doi.org/10.1126/science.1244358} {\bibfield  {journal} {\bibinfo
  {journal} {Science}\ }\textbf {\bibinfo {volume} {342}},\ \bibinfo {pages}
  {614} (\bibinfo {year} {2013})}\BibitemShut {NoStop}%
\bibitem [{\citenamefont {Sandner}\ \emph {et~al.}(2015)\citenamefont
  {Sandner}, \citenamefont {Preis}, \citenamefont {Schell}, \citenamefont
  {Giudici}, \citenamefont {Watanabe}, \citenamefont {Taniguchi}, \citenamefont
  {Weiss},\ and\ \citenamefont {Eroms}}]{Sandner2015}%
  \BibitemOpen
  \bibfield  {author} {\bibinfo {author} {\bibfnamefont {A.}~\bibnamefont
  {Sandner}}, \bibinfo {author} {\bibfnamefont {T.}~\bibnamefont {Preis}},
  \bibinfo {author} {\bibfnamefont {C.}~\bibnamefont {Schell}}, \bibinfo
  {author} {\bibfnamefont {P.}~\bibnamefont {Giudici}}, \bibinfo {author}
  {\bibfnamefont {K.}~\bibnamefont {Watanabe}}, \bibinfo {author}
  {\bibfnamefont {T.}~\bibnamefont {Taniguchi}}, \bibinfo {author}
  {\bibfnamefont {D.}~\bibnamefont {Weiss}},\ and\ \bibinfo {author}
  {\bibfnamefont {J.}~\bibnamefont {Eroms}},\ }\bibfield  {title} {\bibinfo
  {title} {Ballistic transport in graphene antidot lattices},\ }\href
  {https://doi.org/10.1021/acs.nanolett.5b04414} {\bibfield  {journal}
  {\bibinfo  {journal} {Nano Lett.}\ }\textbf {\bibinfo {volume} {15}},\
  \bibinfo {pages} {8402} (\bibinfo {year} {2015})}\BibitemShut {NoStop}%
\bibitem [{\citenamefont {Shalygin}\ \emph {et~al.}(2006)\citenamefont
  {Shalygin}, \citenamefont {Diehl}, \citenamefont {Hoffmann}, \citenamefont
  {Danilov}, \citenamefont {Herrle}, \citenamefont {Tarasenko}, \citenamefont
  {Schuh}, \citenamefont {Gerl}, \citenamefont {Wegscheider}, \citenamefont
  {Prettl},\ and\ \citenamefont {Ganichev}}]{Shalygin2006}%
  \BibitemOpen
  \bibfield  {author} {\bibinfo {author} {\bibfnamefont {V.~A.}\ \bibnamefont
  {Shalygin}}, \bibinfo {author} {\bibfnamefont {H.}~\bibnamefont {Diehl}},
  \bibinfo {author} {\bibfnamefont {C.}~\bibnamefont {Hoffmann}}, \bibinfo
  {author} {\bibfnamefont {S.~N.}\ \bibnamefont {Danilov}}, \bibinfo {author}
  {\bibfnamefont {T.}~\bibnamefont {Herrle}}, \bibinfo {author} {\bibfnamefont
  {S.~A.}\ \bibnamefont {Tarasenko}}, \bibinfo {author} {\bibfnamefont
  {D.}~\bibnamefont {Schuh}}, \bibinfo {author} {\bibfnamefont
  {C.}~\bibnamefont {Gerl}}, \bibinfo {author} {\bibfnamefont {W.}~\bibnamefont
  {Wegscheider}}, \bibinfo {author} {\bibfnamefont {W.}~\bibnamefont
  {Prettl}},\ and\ \bibinfo {author} {\bibfnamefont {S.~D.}\ \bibnamefont
  {Ganichev}},\ }\bibfield  {title} {\bibinfo {title} {Spin photocurrents and
  the circular photon drag effect in (110)-grown quantum well structures},\
  }\href {https://doi.org/10.1134/s0021364006220097} {\bibfield  {journal}
  {\bibinfo  {journal} {JETP Lett.}\ }\textbf {\bibinfo {volume} {84}},\
  \bibinfo {pages} {570} (\bibinfo {year} {2006})}\BibitemShut {NoStop}%
\bibitem [{\citenamefont {Plank}\ \emph {et~al.}(2016)\citenamefont {Plank},
  \citenamefont {Danilov}, \citenamefont {Bel'kov}, \citenamefont {Shalygin},
  \citenamefont {Kampmeier}, \citenamefont {Lanius}, \citenamefont {Mussler},
  \citenamefont {Gr\"utzmacher},\ and\ \citenamefont {Ganichev}}]{Plank2016}%
  \BibitemOpen
  \bibfield  {author} {\bibinfo {author} {\bibfnamefont {H.}~\bibnamefont
  {Plank}}, \bibinfo {author} {\bibfnamefont {S.~N.}\ \bibnamefont {Danilov}},
  \bibinfo {author} {\bibfnamefont {V.~V.}\ \bibnamefont {Bel'kov}}, \bibinfo
  {author} {\bibfnamefont {V.~A.}\ \bibnamefont {Shalygin}}, \bibinfo {author}
  {\bibfnamefont {J.}~\bibnamefont {Kampmeier}}, \bibinfo {author}
  {\bibfnamefont {M.}~\bibnamefont {Lanius}}, \bibinfo {author} {\bibfnamefont
  {G.}~\bibnamefont {Mussler}}, \bibinfo {author} {\bibfnamefont
  {D.}~\bibnamefont {Gr\"utzmacher}},\ and\ \bibinfo {author} {\bibfnamefont
  {S.~D.}\ \bibnamefont {Ganichev}},\ }\bibfield  {title} {\bibinfo {title}
  {Opto-electronic characterization of three dimensional topological
  insulators},\ }\href {https://doi.org/10.1063/1.4965962} {\bibfield
  {journal} {\bibinfo  {journal} {J. Appl. Phys.}\ }\textbf {\bibinfo {volume}
  {120}},\ \bibinfo {pages} {165301} (\bibinfo {year} {2016})}\BibitemShut
  {NoStop}%
\bibitem [{\citenamefont {Dantscher}\ \emph {et~al.}(2017)\citenamefont
  {Dantscher}, \citenamefont {Kozlov}, \citenamefont {Scherr}, \citenamefont
  {Gebert}, \citenamefont {B\"arenf\"anger}, \citenamefont {Durnev},
  \citenamefont {Tarasenko}, \citenamefont {Bel'kov}, \citenamefont
  {Mikhailov}, \citenamefont {Dvoretsky}, \citenamefont {Kvon}, \citenamefont
  {Ziegler}, \citenamefont {Weiss},\ and\ \citenamefont
  {Ganichev}}]{Dantscher2017}%
  \BibitemOpen
  \bibfield  {author} {\bibinfo {author} {\bibfnamefont {K.-M.}\ \bibnamefont
  {Dantscher}}, \bibinfo {author} {\bibfnamefont {D.~A.}\ \bibnamefont
  {Kozlov}}, \bibinfo {author} {\bibfnamefont {M.~T.}\ \bibnamefont {Scherr}},
  \bibinfo {author} {\bibfnamefont {S.}~\bibnamefont {Gebert}}, \bibinfo
  {author} {\bibfnamefont {J.}~\bibnamefont {B\"arenf\"anger}}, \bibinfo
  {author} {\bibfnamefont {M.~V.}\ \bibnamefont {Durnev}}, \bibinfo {author}
  {\bibfnamefont {S.~A.}\ \bibnamefont {Tarasenko}}, \bibinfo {author}
  {\bibfnamefont {V.~V.}\ \bibnamefont {Bel'kov}}, \bibinfo {author}
  {\bibfnamefont {N.~N.}\ \bibnamefont {Mikhailov}}, \bibinfo {author}
  {\bibfnamefont {S.~A.}\ \bibnamefont {Dvoretsky}}, \bibinfo {author}
  {\bibfnamefont {Z.~D.}\ \bibnamefont {Kvon}}, \bibinfo {author}
  {\bibfnamefont {J.}~\bibnamefont {Ziegler}}, \bibinfo {author} {\bibfnamefont
  {D.}~\bibnamefont {Weiss}},\ and\ \bibinfo {author} {\bibfnamefont {S.~D.}\
  \bibnamefont {Ganichev}},\ }\bibfield  {title} {\bibinfo {title}
  {Photogalvanic probing of helical edge channels in two-dimensional {HgTe}
  topological insulators},\ }\href {https://doi.org/10.1103/PhysRevB.95.201103}
  {\bibfield  {journal} {\bibinfo  {journal} {Phys. Rev. B}\ }\textbf {\bibinfo
  {volume} {95}},\ \bibinfo {pages} {201103} (\bibinfo {year}
  {2017})}\BibitemShut {NoStop}%
\bibitem [{\citenamefont {Ganichev}\ \emph {et~al.}(2003)\citenamefont
  {Ganichev}, \citenamefont {Schneider}, \citenamefont {Bel'kov}, \citenamefont
  {Ivchenko}, \citenamefont {Tarasenko}, \citenamefont {Wegscheider},
  \citenamefont {Weiss}, \citenamefont {Schuh}, \citenamefont {Murdin},
  \citenamefont {Phillips}, \citenamefont {Pidgeon}, \citenamefont {Clarke},
  \citenamefont {Merrick}, \citenamefont {Murzyn}, \citenamefont {Beregulin},\
  and\ \citenamefont {Prettl}}]{Ganichev2003}%
  \BibitemOpen
  \bibfield  {author} {\bibinfo {author} {\bibfnamefont {S.~D.}\ \bibnamefont
  {Ganichev}}, \bibinfo {author} {\bibfnamefont {P.}~\bibnamefont {Schneider}},
  \bibinfo {author} {\bibfnamefont {V.~V.}\ \bibnamefont {Bel'kov}}, \bibinfo
  {author} {\bibfnamefont {E.~L.}\ \bibnamefont {Ivchenko}}, \bibinfo {author}
  {\bibfnamefont {S.~A.}\ \bibnamefont {Tarasenko}}, \bibinfo {author}
  {\bibfnamefont {W.}~\bibnamefont {Wegscheider}}, \bibinfo {author}
  {\bibfnamefont {D.}~\bibnamefont {Weiss}}, \bibinfo {author} {\bibfnamefont
  {D.}~\bibnamefont {Schuh}}, \bibinfo {author} {\bibfnamefont {B.~N.}\
  \bibnamefont {Murdin}}, \bibinfo {author} {\bibfnamefont {P.~J.}\
  \bibnamefont {Phillips}}, \bibinfo {author} {\bibfnamefont {C.~R.}\
  \bibnamefont {Pidgeon}}, \bibinfo {author} {\bibfnamefont {D.~G.}\
  \bibnamefont {Clarke}}, \bibinfo {author} {\bibfnamefont {M.}~\bibnamefont
  {Merrick}}, \bibinfo {author} {\bibfnamefont {P.}~\bibnamefont {Murzyn}},
  \bibinfo {author} {\bibfnamefont {E.~V.}\ \bibnamefont {Beregulin}},\ and\
  \bibinfo {author} {\bibfnamefont {W.}~\bibnamefont {Prettl}},\ }\bibfield
  {title} {\bibinfo {title} {Spin-galvanic effect due to optical spin
  orientation in n-type {GaAs} quantum well structures},\ }\href
  {https://doi.org/10.1103/physrevb.68.081302} {\bibfield  {journal} {\bibinfo
  {journal} {Phys. Rev. B}\ }\textbf {\bibinfo {volume} {68}},\ \bibinfo
  {pages} {081302(R)} (\bibinfo {year} {2003})}\BibitemShut {NoStop}%
\bibitem [{\citenamefont {Hubmann}\ \emph {et~al.}(2019)\citenamefont
  {Hubmann}, \citenamefont {Gebert}, \citenamefont {Budkin}, \citenamefont
  {Bel'kov}, \citenamefont {Ivchenko}, \citenamefont {Dmitriev}, \citenamefont
  {Baumann}, \citenamefont {Otteneder}, \citenamefont {Ziegler}, \citenamefont
  {Disterheft}, \citenamefont {Kozlov}, \citenamefont {Mikhailov},
  \citenamefont {Dvoretsky}, \citenamefont {Kvon}, \citenamefont {Weiss},\ and\
  \citenamefont {Ganichev}}]{Hubmann2019}%
  \BibitemOpen
  \bibfield  {author} {\bibinfo {author} {\bibfnamefont {S.}~\bibnamefont
  {Hubmann}}, \bibinfo {author} {\bibfnamefont {S.}~\bibnamefont {Gebert}},
  \bibinfo {author} {\bibfnamefont {G.~V.}\ \bibnamefont {Budkin}}, \bibinfo
  {author} {\bibfnamefont {V.~V.}\ \bibnamefont {Bel'kov}}, \bibinfo {author}
  {\bibfnamefont {E.~L.}\ \bibnamefont {Ivchenko}}, \bibinfo {author}
  {\bibfnamefont {A.~P.}\ \bibnamefont {Dmitriev}}, \bibinfo {author}
  {\bibfnamefont {S.}~\bibnamefont {Baumann}}, \bibinfo {author} {\bibfnamefont
  {M.}~\bibnamefont {Otteneder}}, \bibinfo {author} {\bibfnamefont
  {J.}~\bibnamefont {Ziegler}}, \bibinfo {author} {\bibfnamefont
  {D.}~\bibnamefont {Disterheft}}, \bibinfo {author} {\bibfnamefont {D.~A.}\
  \bibnamefont {Kozlov}}, \bibinfo {author} {\bibfnamefont {N.~N.}\
  \bibnamefont {Mikhailov}}, \bibinfo {author} {\bibfnamefont {S.~A.}\
  \bibnamefont {Dvoretsky}}, \bibinfo {author} {\bibfnamefont {Z.~D.}\
  \bibnamefont {Kvon}}, \bibinfo {author} {\bibfnamefont {D.}~\bibnamefont
  {Weiss}},\ and\ \bibinfo {author} {\bibfnamefont {S.~D.}\ \bibnamefont
  {Ganichev}},\ }\bibfield  {title} {\bibinfo {title} {High-frequency impact
  ionization and nonlinearity of photocurrent induced by intense terahertz
  radiation in {HgTe}-based quantum well structures},\ }\href
  {https://doi.org/10.1103/physrevb.99.085312} {\bibfield  {journal} {\bibinfo
  {journal} {Phys. Rev. B}\ }\textbf {\bibinfo {volume} {99}},\ \bibinfo
  {pages} {085312} (\bibinfo {year} {2019})}\BibitemShut {NoStop}%
\bibitem [{Note1()}]{Note1}%
  \BibitemOpen
  \bibinfo {note} {Note that few final measurements were carried out for
  $V_y^\protect \text {dc}$ applied between contacts neighboring the damaged
  source and drain contacts, which had no apparent effect on the signal
  recorded for the Hall voltage induced in the middle of the Hall
  bar.}\BibitemShut {Stop}%
\bibitem [{Note2()}]{Note2}%
  \BibitemOpen
  \bibinfo {note} {Note that the symmetry in respect to the CNP has been found
  to be sensitive to the cool-down procedure. In experiments differing by the
  cool-down circle only we observed that in several cool downs the signal for
  negative $U_{\protect \rm g}^{\protect \rm eff}$ was almost absent, see
  Fig.~\ref {frequency}(d). This fact together with the mentioned cool-down
  dependent shift of the CNP indicates that the surface charge can be different
  in different measurements and may play an important role in the
  photoconductivity response.}\BibitemShut {Stop}%
\bibitem [{Note3()}]{Note3}%
  \BibitemOpen
  \bibinfo {note} {In Ref.~\protect \rev@citealp {Durnev2021a} the energy
  relaxation rate $\tau _0^{-1}$ is considered to be $\varepsilon
  $-independent.}\BibitemShut {Stop}%
\bibitem [{\citenamefont {Ganichev}\ and\ \citenamefont
  {Prettl}(2005)}]{Ganichev2005}%
  \BibitemOpen
  \bibfield  {author} {\bibinfo {author} {\bibfnamefont {S.~D.}\ \bibnamefont
  {Ganichev}}\ and\ \bibinfo {author} {\bibfnamefont {W.}~\bibnamefont
  {Prettl}},\ }\href
  {https://doi.org/10.1093/acprof:oso/9780198528302.001.0001} {\emph {\bibinfo
  {title} {Intense Terahertz Excitation of Semiconductors}}}\ (\bibinfo
  {publisher} {Oxford University Press},\ \bibinfo {address} {Oxford},\
  \bibinfo {year} {2005})\BibitemShut {NoStop}%
\bibitem [{\citenamefont {Sarkar}\ \emph {et~al.}(2015)\citenamefont {Sarkar},
  \citenamefont {Amin}, \citenamefont {Modak}, \citenamefont {Singh},
  \citenamefont {Mukerjee},\ and\ \citenamefont {Bid}}]{Sarkar2015}%
  \BibitemOpen
  \bibfield  {author} {\bibinfo {author} {\bibfnamefont {S.}~\bibnamefont
  {Sarkar}}, \bibinfo {author} {\bibfnamefont {K.~R.}\ \bibnamefont {Amin}},
  \bibinfo {author} {\bibfnamefont {R.}~\bibnamefont {Modak}}, \bibinfo
  {author} {\bibfnamefont {A.}~\bibnamefont {Singh}}, \bibinfo {author}
  {\bibfnamefont {S.}~\bibnamefont {Mukerjee}},\ and\ \bibinfo {author}
  {\bibfnamefont {A.}~\bibnamefont {Bid}},\ }\bibfield  {title} {\bibinfo
  {title} {Role of different scattering mechanisms on the temperature
  dependence of transport in graphene},\ }\href
  {https://doi.org/10.1038/srep16772} {\bibfield  {journal} {\bibinfo
  {journal} {Sci. Rep.}\ }\textbf {\bibinfo {volume} {5}},\ \bibinfo {pages}
  {16772} (\bibinfo {year} {2015})}\BibitemShut {NoStop}%
\bibitem [{\citenamefont {Ganichev}\ \emph {et~al.}(2002)\citenamefont
  {Ganichev}, \citenamefont {Danilov}, \citenamefont {Bel'kov}, \citenamefont
  {Ivchenko}, \citenamefont {Bichler}, \citenamefont {Wegscheider},
  \citenamefont {Weiss},\ and\ \citenamefont {Prettl}}]{Ganichev2002}%
  \BibitemOpen
  \bibfield  {author} {\bibinfo {author} {\bibfnamefont {S.~D.}\ \bibnamefont
  {Ganichev}}, \bibinfo {author} {\bibfnamefont {S.~N.}\ \bibnamefont
  {Danilov}}, \bibinfo {author} {\bibfnamefont {V.~V.}\ \bibnamefont
  {Bel'kov}}, \bibinfo {author} {\bibfnamefont {E.~L.}\ \bibnamefont
  {Ivchenko}}, \bibinfo {author} {\bibfnamefont {M.}~\bibnamefont {Bichler}},
  \bibinfo {author} {\bibfnamefont {W.}~\bibnamefont {Wegscheider}}, \bibinfo
  {author} {\bibfnamefont {D.}~\bibnamefont {Weiss}},\ and\ \bibinfo {author}
  {\bibfnamefont {W.}~\bibnamefont {Prettl}},\ }\bibfield  {title} {\bibinfo
  {title} {Spin-sensitive bleaching and monopolar spin orientation in quantum
  wells},\ }\href {https://doi.org/10.1103/physrevlett.88.057401} {\bibfield
  {journal} {\bibinfo  {journal} {Phys. Rev. Lett.}\ }\textbf {\bibinfo
  {volume} {88}},\ \bibinfo {pages} {057401} (\bibinfo {year}
  {2002})}\BibitemShut {NoStop}%
\bibitem [{\citenamefont {Danilov}\ \emph {et~al.}(2021)\citenamefont
  {Danilov}, \citenamefont {Golub}, \citenamefont {Mayer}, \citenamefont
  {Beer}, \citenamefont {Binder}, \citenamefont {Mönch}, \citenamefont
  {Min{\'{a}}r}, \citenamefont {Kronseder}, \citenamefont {Back}, \citenamefont
  {Bougeard},\ and\ \citenamefont {Ganichev}}]{Danilov2021}%
  \BibitemOpen
  \bibfield  {author} {\bibinfo {author} {\bibfnamefont {S.~N.}\ \bibnamefont
  {Danilov}}, \bibinfo {author} {\bibfnamefont {L.~E.}\ \bibnamefont {Golub}},
  \bibinfo {author} {\bibfnamefont {T.}~\bibnamefont {Mayer}}, \bibinfo
  {author} {\bibfnamefont {A.}~\bibnamefont {Beer}}, \bibinfo {author}
  {\bibfnamefont {S.}~\bibnamefont {Binder}}, \bibinfo {author} {\bibfnamefont
  {E.}~\bibnamefont {Mönch}}, \bibinfo {author} {\bibfnamefont
  {J.}~\bibnamefont {Min{\'{a}}r}}, \bibinfo {author} {\bibfnamefont
  {M.}~\bibnamefont {Kronseder}}, \bibinfo {author} {\bibfnamefont {C.~H.}\
  \bibnamefont {Back}}, \bibinfo {author} {\bibfnamefont {D.}~\bibnamefont
  {Bougeard}},\ and\ \bibinfo {author} {\bibfnamefont {S.~D.}\ \bibnamefont
  {Ganichev}},\ }\bibfield  {title} {\bibinfo {title} {Superlinear
  photogalvanic effects in $( \textrm{Bi}_0.3\textrm{Sb}_0.7)_2 (
  \textrm{Te}_0.1\textrm{Se}_0.9)_3$ : Probing three-dimensional topological
  insulator surface states at room temperature},\ }\href
  {https://doi.org/10.1103/physrevapplied.16.064030} {\bibfield  {journal}
  {\bibinfo  {journal} {Phys. Rev. Appl}\ }\textbf {\bibinfo {volume} {16}},\
  \bibinfo {pages} {064030} (\bibinfo {year} {2021})}\BibitemShut {NoStop}%
\bibitem [{\citenamefont {Mics}\ \emph {et~al.}(2015)\citenamefont {Mics},
  \citenamefont {Tielrooij}, \citenamefont {Parvez}, \citenamefont {Jensen},
  \citenamefont {Ivanov}, \citenamefont {Feng}, \citenamefont {Müllen},
  \citenamefont {Bonn},\ and\ \citenamefont {Turchinovich}}]{Mics2015}%
  \BibitemOpen
  \bibfield  {author} {\bibinfo {author} {\bibfnamefont {Z.}~\bibnamefont
  {Mics}}, \bibinfo {author} {\bibfnamefont {K.-J.}\ \bibnamefont {Tielrooij}},
  \bibinfo {author} {\bibfnamefont {K.}~\bibnamefont {Parvez}}, \bibinfo
  {author} {\bibfnamefont {S.~A.}\ \bibnamefont {Jensen}}, \bibinfo {author}
  {\bibfnamefont {I.}~\bibnamefont {Ivanov}}, \bibinfo {author} {\bibfnamefont
  {X.}~\bibnamefont {Feng}}, \bibinfo {author} {\bibfnamefont {K.}~\bibnamefont
  {Müllen}}, \bibinfo {author} {\bibfnamefont {M.}~\bibnamefont {Bonn}},\ and\
  \bibinfo {author} {\bibfnamefont {D.}~\bibnamefont {Turchinovich}},\
  }\bibfield  {title} {\bibinfo {title} {Thermodynamic picture of ultrafast
  charge transport in graphene},\ }\href {https://doi.org/10.1038/ncomms8655}
  {\bibfield  {journal} {\bibinfo  {journal} {Nat. Commun.}\ }\textbf {\bibinfo
  {volume} {6}},\ \bibinfo {pages} {7655} (\bibinfo {year} {2015})}\BibitemShut
  {NoStop}%
\end{thebibliography}%

\end{document}